\documentclass[12pt]{article}

\usepackage[a4paper,text={16.8cm,22.4cm}]{geometry}
\usepackage{amsmath,amsfonts,slashed,amssymb,tikz,bm,psfrag,graphicx,color,dsfont}
\usepackage{multicol}

\RequirePackage[sort&compress,square,comma,numbers]{natbib}
\allowdisplaybreaks 
\begin{document}

\begin{titlepage}

\begin{flushright}
\normalsize
MITP/15-031\\
May 14, 2015
\end{flushright}

\vspace{0.1cm}
\begin{center}
\Large\bf
Exclusive Radiative Higgs Decays as Probes of Light-Quark Yukawa Couplings 
\end{center}

\vspace{0.5cm}
\begin{center}
Matthias K\"onig$^a$ and Matthias Neubert$^{a,b,c}$\\
\vspace{0.7cm} 
{\sl ${}^a$PRISMA Cluster of Excellence \& Mainz Institute for Theoretical Physics\\
Johannes Gutenberg University, 55099 Mainz, Germany\\[3mm]
${}^b$Institut f\"ur Theoretische Physik\\ 
Universit\"at Heidelberg, Philosophenweg 16, 69120 Heidelberg, Germany\\[3mm]
${}^c$Department of Physics, LEPP, Cornell University, Ithaca, NY 14853, U.S.A.}
\end{center}

\vspace{0.2cm}
\begin{abstract}
We present a detailed analysis of the rare exclusive Higgs-boson decays into a single vector meson and a photon and investigate the possibility of using these processes to probe the light-quark Yukawa couplings. We work with an effective Lagrangian with modified Higgs couplings to account for possible new-physics effects in a model-independent way. The $h\to V\gamma$ decay rate is governed by the destructive interference of two amplitudes, one of which involves the Higgs coupling to the quark anti-quark pair inside the vector meson. We derive this amplitude at next-to-leading order in $\alpha_s$ using QCD factorization, including the resummation of large logarithmic corrections and accounting for the effects of flavor mixing. The high factorization scale $\mu\sim m_h$ ensures that our results are rather insensitive to the hadronic parameters characterizing the light-cone distribution amplitude of the vector meson. The second amplitude arises from the loop-induced effective $h\gamma\gamma^*$ and $h\gamma Z^*$ couplings, where the off-shell gauge boson converts into the vector meson. We devise a strategy to eliminate theoretical uncertainties related to this amplitude to almost arbitrary precision. This opens up the possibility to probe for ${\cal O}(1)$ modifications of the $c$- and $b$-quark Yukawa couplings and ${\cal O}(30)$ modifications of the $s$-quark Yukawa coupling in the high-luminosity LHC run. In particular, we show that measurements of the ratios $\mbox{Br}(h\to\Upsilon(nS)\,\gamma)/\mbox{Br}(h\to\gamma\gamma)$ and $\mbox{Br}(h\to b\bar b)/\mbox{Br}(h\to\gamma\gamma)$ can provide complementary information on the real and imaginary parts of the $b$-quark Yukawa coupling. More accurate measurements would be possible at a future 100\,TeV proton-proton collider.
\end{abstract}
\vfil

\end{titlepage}

\section{Introduction}

The discovery of the Higgs boson in 2012 \cite{Aad:2012tfa,Chatrchyan:2012ufa} has established the existence of a new kind of elementary particle, which couples to the other particles of the Standard Model (SM) in a non-universal way. The SM predictions that the Higgs couplings to heavy gauge bosons and fermions are given by $2m_{W,Z}^2/v$ and $m_f/v$, where $v\approx 246$\,GeV is the Higgs vacuum expectation value, have been confirmed within experimental uncertainties for the $W$ and $Z$ bosons and for the third-generation fermions \cite{Khachatryan:2014jba,ATLAS_Higgs}. However, no direct measurements of the Higgs couplings to the light fermions of the first two generations are available at present. It is not difficult to come up with models in which these couplings can deviate significantly from those predicted in the SM. For example, in \cite{Giudice:2008uua,Bauer:2015fxa} it was proposed that the Yukawa couplings may depend in a non-trivial way on the Higgs field, and that this might explain the hierarchies seen in the spectrum of fermion masses. A more general analysis of different classes of models in which the Higgs couplings to fermions can differ significantly from those of the SM was presented in \cite{Harnik:2012pb}. Probing the Higgs couplings to light fermions is thus of paramount importance. This includes both flavor-diagonal and flavor-changing interactions. Correlations between the two types of couplings, which to some extent are model dependent, have been studied in \cite{Goertz:2014qia}. The CMS collaboration has recently reported a slight excess in the search for the flavor-violating decay $h\to\tau^\pm\mu^\mp$ \cite{CMS:2014hha}, which, if interpreted as a signal, corresponds to a branching fraction $\mbox{Br}(h\to\tau^\pm\mu^\mp)=(0.89\,_{-\,0.37}^{+\,0.40})\,\%$. Not surprisingly, this observation has led to much theoretical speculation.

Our focus in the present work is on the Higgs couplings to light quarks ($q\ne t$). In a couple of beautiful papers, it has recently been proposed that one might get access to these couplings by focussing on the rare, exclusive decays $h\to V\gamma$ of the Higgs boson \cite{Bodwin:2013gca,Kagan:2014ila}, where the final state contains a single vector meson $V$. Such measurements are extremely challenging at the LHC, as the corresponding branching fractions are in the range of few times $10^{-6}$. Nevertheless, observing these processes is not hopeless in view of the fact that in its high-luminosity run the LHC will serve as a Higgs factory. With 3\,ab$^{-1}$ of integrated luminosity, about $1.7\times 10^8$ Higgs bosons per experiment will have been produced \cite{Dawson:2013bba}, and even larger Higgs samples could be obtained at a future facility such as a 100\,TeV proton-proton collider. The theoretical description of rare exclusive decays employs the formalism of QCD factorization \cite{Lepage:1979zb,Lepage:1980fj,Efremov:1978rn,Efremov:1979qk,Chernyak:1983ej}, which was originally developed for the analysis of hard exclusive QCD processes and later extended to the more complicated case of exclusive hadronic two-body decays of $B$ mesons \cite{Beneke:1999br,Beneke:2000ry}. In a recent paper, we have systematically developed this approach for the case of the exclusive decays $Z\to M\gamma$, $W\to M\gamma$ and $Z\to MW$ \cite{Grossman:2015lea}. These processes are less susceptible to new physics and can therefore be used to test the QCD factorization approach and extract valuable information about the light-cone distribution amplitudes (LCDAs) of various mesons $M$. Observing exclusive radiative decays of heavy electroweak gauge bosons in the high-luminosity run of the LHC would provide a proof of principle that this kind of rare-decay searches can be performed in a hadron-collider environment. A promising first step in the direction of observing the decays $Z\to J/\psi\,\gamma$ and $Z\to\Upsilon\gamma$, along with the corresponding Higgs decays $h\to J/\psi\,\gamma$ and $h\to\Upsilon\gamma$, has recently been reported by the ATLAS collaboration~\cite{Aad:2015sda}. 

In this work we extend previous studies of exclusive radiative decays of the Higgs boson in several important ways. We include QCD radiative corrections and resum large logarithms arising from evolution effects between the Higgs mass scale and a low hadronic scale. We comment on the structure of power-suppressed corrections to the factorization formula, study the effects of the off-shellness of the photon in the $h\to\gamma\gamma^*\to\gamma V$ process and include the power-suppressed $h\to\gamma Z^*\to \gamma V$ contribution. We also take into account the effects of $\omega\!-\!\phi$ mixing. Most importantly, our analysis allows for generic non-standard Higgs couplings to SM fermions and gauge bosons, including CP-odd interactions. We devise a strategy which allows us to eliminate the theoretical uncertainties related to the dominant $h\to\gamma\gamma^*\to\gamma V$ conversion contribution, including possible new-physics effects, to almost arbitrary precision. This is a crucial prerequisite for achieving the desired sensitivity to the Yukawa couplings of light quarks. Finally, we address all relevant $h\to V\gamma$ decays to both light and heavy mesons in one coherent formalism. For technical details on the QCD factorization approach the reader is referred to \cite{Grossman:2015lea}.

In the following section we start by defining effective couplings of the Higgs boson to SM fermions and gauge bosons. These include both CP-even and CP-odd couplings. In Section~\ref{sec:radiativedec} we discuss in detail the calculation of the different contributions to the $h\to V\gamma$ decay amplitudes in the QCD factorization approach, distinguishing between ``direct'' contributions induced by the Yukawa couplings of the Higgs boson to light quarks ($q\ne t$) and ``indirect contributions'' resulting from a $h\to\gamma\gamma^*/\gamma Z^*$ transition followed by the conversion of the off-shell boson into a vector meson. We show that uncertainties related to the effective $h\gamma\gamma$ coupling strength, as regards to both theoretical uncertainties and possible new-physics contributions, can be eliminated by studying the ratio of the $h\to V\gamma$ and $h\to\gamma\gamma$ branching fractions. In Section~\ref{sec:pheno} we present a phenomenological analysis both in the SM and in generic new-physics models with modified Higgs interactions. In particular, we point out that measurements of the ratios of the $h\to\Upsilon(nS)\,\gamma$, $h\to b\bar b$ and $h\to\gamma\gamma$ branching fractions can yield highly complementary information on the real and imaginary parts of the bottom-quark Yukawa coupling. Section~\ref{sec:concl} contains a summary of our results and the conclusions. Technical details are relegated to five appendices.

\section{Effective Higgs couplings}
\label{sec:Leff}

In our analysis we assume SM couplings for all particles other than the Higgs boson. For the Higgs couplings to SM quarks and gauge bosons we adopt the effective Lagrangian 
\begin{equation}\label{eqn:effL}
\begin{aligned}
   {\cal L}_{\rm eff}^{\rm Higgs} &= \kappa_W\,\frac{2m_W^2}{v}\,h\,W_\mu^+ W^{-\mu} 
    + \kappa_Z\,\frac{m_Z^2}{v}\,h\,Z_\mu Z^\mu 
    - \sum_f\,\frac{m_f}{v}\,h\,\bar f \left( \kappa_f + i\tilde\kappa_f\gamma_5 \right) f \\[-3mm]
   &\quad\mbox{}+ \frac{\alpha}{4\pi v} \left( 
    \kappa_{\gamma\gamma}\,h\,F_{\mu\nu} F^{\mu\nu}
    - \tilde\kappa_{\gamma\gamma}\,h\,F_{\mu\nu} \tilde F^{\mu\nu} 
    + \frac{2\kappa_{\gamma Z}}{s_W c_W}\,h\,F_{\mu\nu} Z^{\mu\nu} 
    - \frac{2\tilde\kappa_{\gamma Z}}{s_W c_W}\,h\,F_{\mu\nu} \tilde Z^{\mu\nu} \right)
    + \dots \,,
\end{aligned}
\end{equation}
where $s_W\equiv\sin\theta_W$ and $c_W\equiv\cos\theta_W$ are the sine and cosine of the weak mixing angle. We use $s_W^2=0.23126\pm 0.00005$ as determined from the neutral-current couplings of the $Z$ boson evaluated at $\mu=m_Z$ \cite{Agashe:2014kda}. $\tilde F^{\mu\nu}=\frac12\epsilon^{\mu\nu\alpha\beta} F_{\alpha\beta}$ is the dual field-strength tensor, and we use a sign convention where $\epsilon^{0123}=1$. Our choice of factoring out a loop factor in the second line is made for later convenience. For SM extensions in which the new particles are heavy, the coefficients of these higher-dimensional operators are suppressed by two powers of the new-physics scale. We emphasize that the above is not a complete list of operators. For instance, we have not included higher-dimensional operators of the form $h W_{\mu\nu} W^{\mu\nu}$ and $h W_{\mu\nu}\tilde W^{\mu\nu}$, whose coefficients are already strongly constrained by data. These operators would enter our analysis only via the $h\to\gamma\gamma^*/\gamma Z^*$ one-loop amplitudes, and without loss of generality their effects can be absorbed into the coefficients $\kappa_{\gamma\gamma}$, $\kappa_{\gamma Z}$ and $\tilde\kappa_{\gamma\gamma}$, $\tilde\kappa_{\gamma Z}$. 

Both the CP-even couplings $\kappa_i$ and the CP-odd coefficients $\tilde\kappa_i$ are real. In the SM $\kappa_W=\kappa_Z=\kappa_f=1$, while $\tilde\kappa_f$ and all the remaining couplings in the second line vanish. Our $\kappa_q$ and $\tilde\kappa_q$ parameters for quarks are related to the corresponding Yukawa couplings by
\begin{equation}\label{kappaq}
   \frac{y_q}{\sqrt2}\equiv (\kappa_q+i\tilde\kappa_q)\,\frac{m_q}{v}
   \equiv (\bar\kappa_q+i\bar{\tilde\kappa}_q)\,\frac{m_b}{v} \,.
\end{equation}
In the last step we have introduced rescaled parameters normalized to the mass of the $b$ quark. This will turn out to be a useful definition for the quarks of the first two generations. 

For our analysis we need the quark masses at the scale $\mu=m_h$, where $m_h=(125.09\pm 0.24)$\,GeV is the Higgs-boson mass \cite{Aad:2015zhl}. We define the running quark masses at next-to-next-to-leading order (NNLO) in the $\overline{\rm MS}$ scheme, starting from the low-energy values given in \cite{Agashe:2014kda}. This yields $m_b(m_h)=(2.79\pm 0.02)$\,GeV, $m_c(m_h)=(622\pm 12)$\,MeV, $m_s(m_h)=(52.8\pm 1.4)$\,MeV, $m_d(m_h)=(2.66\pm 0.11)$\,MeV, and $m_u(m_h)=(1.21\pm 0.08)$\,MeV. For the top quark we obtain $m_t(m_h)=(166.8\pm 0.7)$\,GeV starting from $m_t(m_t)=(163.4\pm 0.7)$\,GeV, which we have derived from the present world average obtained in \cite{ATLAS:2014wva} using the conversion tables provided in \cite{Moch:2014tta}. Present LHC data are largely insensitive to the Yukawa couplings to light quark flavors. From a global $\chi^2$ fit to the measured Higgs rates, the authors of \cite{Kagan:2014ila,Perez:2015aoa} have derived the bounds $\sqrt{|\bar\kappa_u|^2+|\bar{\tilde\kappa}_u|^2}<1.3$ and $\sqrt{|\bar\kappa_{d,s,c}|^2+|\bar{\tilde\kappa}_{d,s,c}|^2}<1.4$ at 95\% confidence level (CL). The corresponding bounds for the original parameters are $\sqrt{|\kappa_u|^2+|\tilde\kappa_u|^2}<3000$, $\sqrt{|\kappa_d|^2+|\tilde\kappa_d|^2}<1500$, $\sqrt{|\kappa_s|^2+|\tilde\kappa_s|^2}<75$ and $\sqrt{|\kappa_c|^2+|\tilde\kappa_c|^2}<6.2$. 

\begin{figure}
\begin{center}
\includegraphics[width=0.22\textwidth]{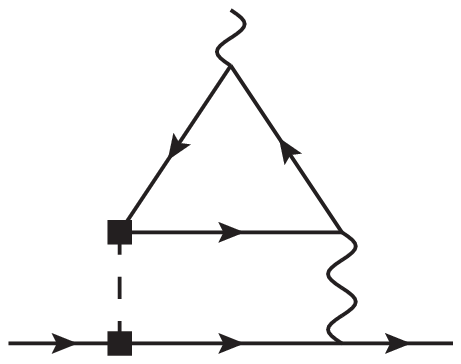} \quad
\includegraphics[width=0.22\textwidth]{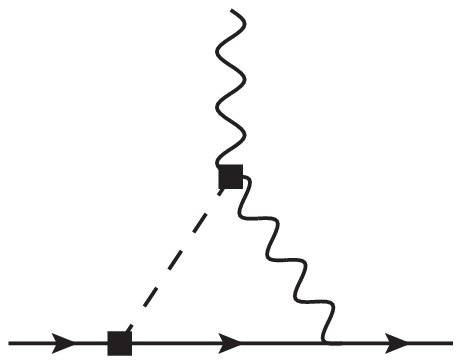}
\parbox{15.5cm}
{\caption{\label{fig:eEDM}
Two-loo Barr-Zee diagram (left) and effective one-loop contribution (right) to the EDM of the electron arising from the CP-odd couplings $\tilde\kappa_f$, $\tilde\kappa_{\gamma\gamma}$ and $\tilde\kappa_{\gamma Z}$ in the effective Lagrangian (\ref{eqn:effL}).}}
\end{center}
\end{figure}

Bounds on the CP-violating Higgs couplings to third-generation fermions have been studied in \cite{Brod:2013cka}. Under the assumption that the Higgs couples to the electron in the standard way ($\kappa_e=1$, $\tilde\kappa_e=0$), the strongest constraints come from the electric dipole moment (EDM) of the electron. They arise from two-loop Barr-Zee diagrams such as the one shown on the left in Figure~\ref{fig:eEDM} and yield $|\tilde\kappa_t|<0.01$, $|\tilde\kappa_b|<1.9$ and $|\tilde\kappa_\tau|<2.4$ at 90\% CL. Bounds from the neutron EDM are approximately one order of magnitude weaker, but they do not reply on assumptions about the Higgs couplings to the electron. The $h\to b\bar b$ and $h\to\tau^+\tau^-$ rate measurements at the LHC can be used to place upper limits on the combinations $|\kappa_{b,\tau}|^2+|\tilde\kappa_{b,\tau}|^2$, which imply the stronger bounds (at 95\% CL) $|\tilde\kappa_b|<1.44$ and $|\tilde\kappa_\tau|<1.24$ from CMS \cite{Khachatryan:2014jba}, and $|\tilde\kappa_b|<1.3$ and $|\tilde\kappa_\tau|<1.5$ from ATLAS \cite{ATLAS_Higgs}. Bounds from the electron EDM can also be derived for the local operators multiplied by $\tilde\kappa_{\gamma\gamma}$ and $\tilde\kappa_{\gamma Z}$ in (\ref{eqn:effL}). Evaluating the one-loop contributions to the electron EDM shown on the right in Figure~\ref{fig:eEDM}, we obtain in the $\overline{\rm MS}$ scheme
\begin{equation}\label{derela}
\begin{aligned}
   \frac{d_e}{e} 
   &= - \frac{\alpha}{16\pi^3}\,\frac{m_e}{v^2}\,\bigg[
    \left( \ln\frac{\mu^2}{m_h^2} + \frac32 \right)
    \left( \tilde\kappa_{\gamma\gamma}\,\kappa_e
    + \kappa_{\gamma\gamma}\,\tilde\kappa_e \right) \\
   &\hspace{2.5cm}\mbox{}+ \frac{1-4 s_W^2}{4s_W^2 c_W^2}
    \left( \ln\frac{\mu^2}{m_h^2} + \frac32 + \frac{x_Z\ln x_Z}{1-x_Z} \right)
    \left( \tilde\kappa_{\gamma Z}\,\kappa_e
    + \kappa_{\gamma Z}\,\tilde\kappa_e \right) \bigg] \,,
\end{aligned}
\end{equation}
where $x_Z=m_Z^2/m_h^2$. This contribution is logarithmically UV divergent, because the inner structure of the effective $h\gamma\gamma$ and $h\gamma Z$ vertices is not resolved. The term proportional to $\tilde\kappa_{\gamma\gamma}\,\kappa_e$ agrees with a calculation performed in \cite{McKeen:2012av}. The subtraction scale $\mu$ should be identified with the scale of the new physics, which is responsible for these non-standard interactions. Setting $\mu=\Lambda_{\rm NP}=1$\,TeV for an estimate, and assuming SM-like Higgs couplings to the electron, we obtain from the present experimental bound $|d_e|<8.7\cdot 10^{-29}\,e$\,cm (at 90\% CL) \cite{Baron:2013eja} the constraint
\begin{equation}
   \big| \tilde\kappa_{\gamma\gamma} + 0.09\,\tilde\kappa_{\gamma Z} \big|
   < 0.006 \quad \mbox{(90\% CL)} \,.
\end{equation}
Barring a fine tuning of the two contributions, this implies that $|\tilde\kappa_{\gamma\gamma}|<6\cdot 10^{-3}$ and $|\tilde\kappa_{\gamma Z}|<0.07$. If the new-physics scale lies above 1\,TeV then these bounds become stronger. With $\Lambda_{\rm NP}=10$\,TeV, for example, they improve by a factor of~2.

\section{Radiative hadronic decays of Higgs bosons}
\label{sec:radiativedec}

Our focus in this work is on the rare, exclusive radiative decays $h\to V\gamma$, where $V$ denotes a vector meson with momentum $k$ and the photon carries momentum $q$. We will refer to vectors orthogonal to the plane spanned by $k$ and $q$ as being transverse. Up to tiny corrections of order $(m_V/m_h)^2$, the mass of the vector meson can be set to zero. 

\begin{figure}
\begin{center}
\includegraphics[width=0.28\textwidth]{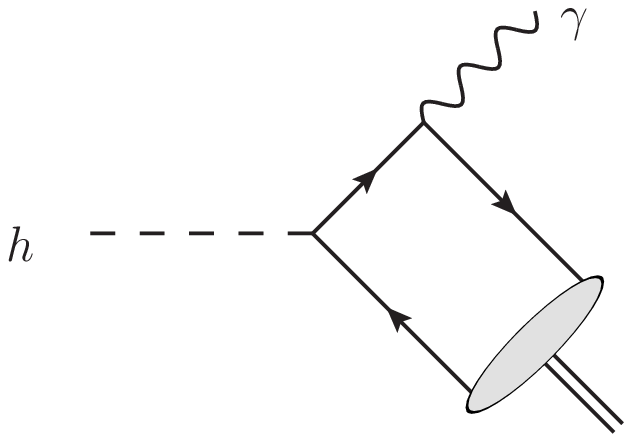}
\includegraphics[width=0.28\textwidth]{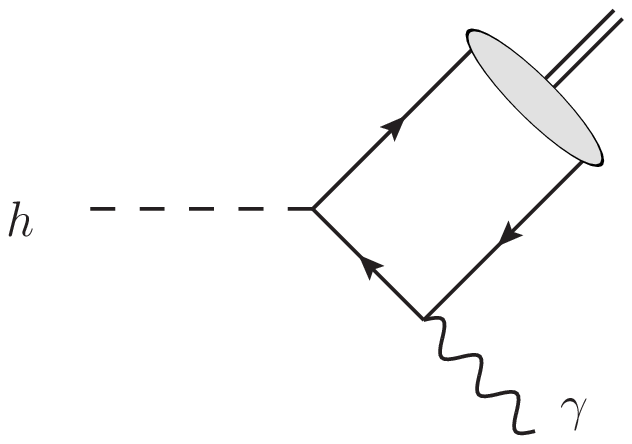}
\includegraphics[width=0.25\textwidth]{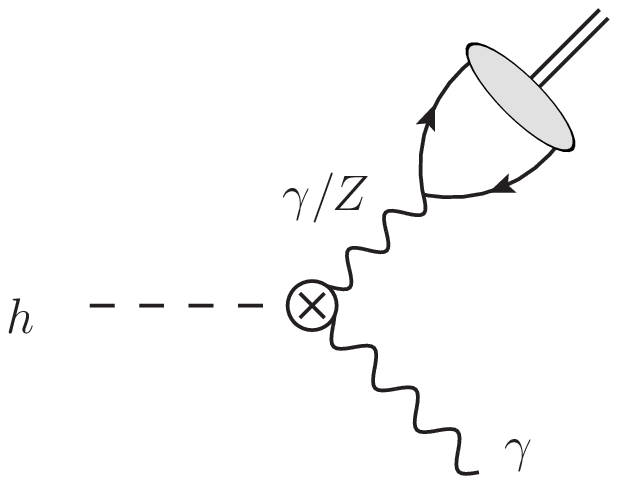}
\parbox{15.5cm}
{\caption{\label{fig:diags}
Direct (left and center) and indirect (right) contributions to the $h\to V\gamma$ decay amplitude. The crossed circle in the third diagram denotes the off-shell $h\to\gamma\gamma^*$ and $h\to\gamma Z^*$ amplitudes, which in the SM arise first at one-loop order.}}
\end{center}
\end{figure}

The leading-order Feynman diagrams are shown in Figure~\ref{fig:diags}. In the first two graphs, the Higgs boson couples to the quark and anti-quark pair from which the meson is formed. We refer to this as the ``direct'' contribution to the decay amplitude \cite{Shifman:1980dk,Keung:1983ac}. It can be calculated from first principles using the QCD factorization approach \cite{Lepage:1979zb,Lepage:1980fj,Efremov:1978rn,Efremov:1979qk,Chernyak:1983ej}, because the energy released to the final-state meson is much larger than the scale of long-distance hadronic physics \cite{Wang:2013ywc,Kagan:2014ila,Bodwin:2014bpa}. At leading power in an expansion in $\Lambda_{\rm QCD}/m_h$, the direct contribution can be expressed as a convolution of a calculable hard-scattering coefficient with the leading-twist LCDA of the vector meson $V$. The corresponding factorization formula was derived in \cite{Grossman:2015lea} using the formalism of soft-collinear effective theory \cite{Bauer:2000yr,Bauer:2001yt,Bauer:2002nz,Beneke:2002ph}. It was shown that, for a given helicity amplitude, the power corrections to the leading term are suppressed by $(\Lambda_{\rm QCD}/m_h)^2$ for light mesons and $(m_Q/m_h)^2$ for mesons containing heavy quarks of flavor $Q$. Even for the $b$-quark these power corrections are negligible. The third diagram in Figure~\ref{fig:diags} shows a different production mechanism, in which the vector meson is produced via the conversion of an off-shell photon or $Z$ boson produced in a $h\to\gamma\gamma^*/\gamma Z^*$ transition \cite{Bodwin:2013gca}. We refer to this as the ``indirect'' contribution. It involves the hadronic matrix element of a local current and thus can be expressed in terms of the decay constant $f_V$ of the vector meson. The direct contribution is sensitive to the Yukawa coupling of the Higgs boson to the quarks which make up the vector meson. We shall find that in the SM the direct and indirect contributions to the $h\to V\gamma$ decay amplitude interfere destructively. They are of similar size for $V=\Upsilon$, while the direct contributions are smaller than the indirect ones by factors of about 0.06 for $V=J/\psi$, 0.002 for $V=\phi$, and few times $10^{-5}$ for $V=\rho^0$ and $\omega$. The sensitivity to the Yukawa couplings thus crucially relies on the precision with which the indirect contributions can be calculated. We will come back to this point below.

The most general parametrization of the $h\to V\gamma$ decay amplitude is
\begin{equation}\label{Ampl}
   i{\cal A}(h\to V\gamma) 
   = - \frac{e f_V}{2} \left[ \left( \varepsilon_V^*\cdot\varepsilon_\gamma^*
    - \frac{q\cdot\varepsilon_V^*\,k\cdot\varepsilon_\gamma^*}{k\cdot q} \right) F_1^V 
    - i\epsilon_{\mu\nu\alpha\beta}\,
    \frac{k^\mu q^\nu\varepsilon_V^{*\alpha}\varepsilon_\gamma^{*\beta}}{k\cdot q}\,F_2^V \right] ,
\end{equation}
where both the final-state meson and the photon are transversely polarized. From (\ref{Ampl}), the decay rate is obtained as
\begin{equation}\label{hVgarate}
   \Gamma(h\to V\gamma) = \frac{\alpha f_V^2}{8 m_h}
    \left( \left| F_1^V \right|^2 + \left| F_2^V \right|^2 \right) .
\end{equation}
Here $\alpha=1/137.036$ is the fine-structure constant evaluated at $q^2=0$ \cite{Agashe:2014kda}, as appropriate for a real photon. We choose to normalize the decay amplitude in (\ref{Ampl}) to the vector-meson decay constant $f_V$, which is defined in terms of a matrix element of a local vector current. Since we consider neutral, flavor-diagonal mesons, the definition of the decay constants (and of other hadronic matrix elements) is complicated by the effects of flavor mixing. In complete generality, such a neutral meson $V$ can be regarded as a superposition of flavor states $|q\bar q\rangle$. We can thus define flavor-dependent decay constants $f_V^q$ via
\begin{equation}\label{fVqdef}
   \langle V(k,\varepsilon)|\,\bar q\gamma^\mu q\,|0\rangle 
   = -if_V^q m_V \varepsilon^{*\mu} \,; \quad q=u,d,s,\dots \,.
\end{equation}
A certain combination of these flavor-specific decay constants can be measured in the leptonic decay $V\to e^+ e^-$. The corresponding decay amplitude involves the matrix element of the electromagnetic current $J_{\rm em}^\mu=\sum_q Q_q\,\bar q\gamma^\mu q$. We thus define
\begin{equation}\label{fVdef}
   Q_V f_V \equiv \sum_q Q_q f_V^q \,, \qquad \mbox{where} \quad
   Q_V = \sum_q c_q^V Q_q \,. 
\end{equation}
Here $c_q^V$ denote the flavor coefficients in the naive constituent-quark model, where $|\rho^0\rangle=\frac{1}{\sqrt2}\left(|u\bar u\rangle-|d\bar d\rangle\right)$, $|\omega\rangle=\frac{1}{\sqrt2}\left(|u\bar u\rangle+|d\bar d\rangle\right)$, $|\phi\rangle=|s\bar s\rangle$ etc. It follows that
\begin{equation}
   \frac{1}{\sqrt2}\,f_{\rho^0} = \sum_q Q_q\,f_{\rho^0}^q \,, \qquad
   \frac{1}{3\sqrt2}\,f_{\omega} = \sum_q Q_q\,f_{\omega}^q \,, \qquad
   - \frac{1}{3}\,f_{\phi} = \sum_q Q_q\,f_{\phi}^q \,, 
\end{equation}
and so on. With these definitions, the electromagnetic decay rate is given by
\begin{equation}\label{GamVee}
   \Gamma(V\to e^+ e^-) = \frac{4\pi Q_V^2 f_V^2}{3m_V}\,\alpha^2(m_V) \,.
\end{equation}
In \cite{Grossman:2015lea}, the so-defined decay constants $f_V$ have been extracted using the most recent experimental data. Our motivation for factoring out these decay constants in (\ref{Ampl}) is that they can be determined experimentally without adopting any specific model for flavor mixing. As we will discuss in Section~\ref{subsec:indirect}, the dominant terms in the indirect contributions to the $h\to V\gamma$ decay amplitudes are proportional to precisely these quantities.

\subsection{Direct contributions to the form factors}

We first consider the direct contribution to the decay amplitude shown in the first two diagrams in Figure~\ref{fig:diags}, in which the Higgs boson couples to the quarks contained inside the vector meson. In order to calculate the corresponding contributions to the form factors $F_i^V$ defined in (\ref{Ampl}), one calculates the corresponding partonic amplitudes with on-shell quark and anti-quark states and then projects these amplitudes onto the leading-twist LCDA of a transversely polarized vector meson $V$. The relevant light-cone projector reads \cite{Beneke:2000wa}
\begin{equation}\label{LCDAV}
   M_{V_\perp}(k,x,\mu) = \frac{if_V^\perp(\mu)}{4}\,\rlap{\hspace{0.1mm}/}{k}\,
    \rlap/\varepsilon_V^{\perp *} \phi_V^\perp(x,\mu) + \dots \,,
\end{equation}
where the dots stand for higher-twist contributions. In analogy with (\ref{fVqdef}) and (\ref{fVdef}), we define a set of flavor-specific transverse decay constants via
\begin{equation}\label{vectorme}
   \langle V(k,\varepsilon)|\,\bar q\,i\sigma^{\mu\nu} q\,|0\rangle 
   = if_V^{q\perp}(\mu) \left( k^\mu \varepsilon^{*\nu} - k^\nu\varepsilon^{*\mu} \right) .
\end{equation}
The quantity $f_V^\perp$ entering (\ref{LCDAV}) is then defined by the combination $f_V^\perp\equiv\big(\sum_q Q_q f_V^{q\perp}\big)/Q_V$. The transverse decay constants are scale-dependent quantities, since the QCD tensor current has a non-zero anomalous dimension. The leading-twist LCDA $\phi_V^\perp(x,\mu)$ can be interpreted as the amplitude for finding a quark with longitudinal momentum fraction $x$ inside the meson.\footnote{Strictly speaking, one should introduce flavor-specific LCDAs $\phi_V^{q\perp}(x,\mu)$ and define  the product $f_V^{q\perp}(\mu)\,\phi_V^{q\perp}(x,\mu)$ in terms of a matrix element of a non-local quark current with flavor $q$, in analogy with relation (4) in \cite{Grossman:2015lea}. Because the LCDAs are normalized to~1, and given the present large uncertainties in the shapes of these functions, it is a safe approximation to employ SU(3) symmetry and replace $\phi_V^{q\perp}(x,\mu)\to\phi_V^\perp(x,\mu)$.} 
It depends on the choice of the factorization scale $\mu$ employed in the factorization formula.

\begin{figure}
\begin{center}
\includegraphics[width=0.21\textwidth]{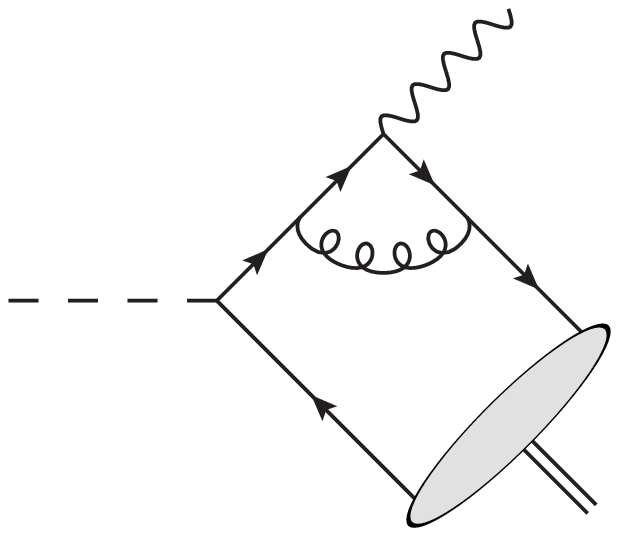}
\includegraphics[width=0.21\textwidth]{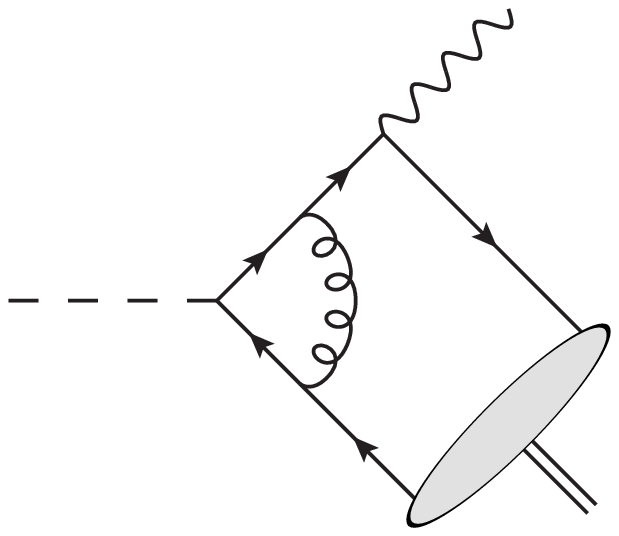}
\includegraphics[width=0.21\textwidth]{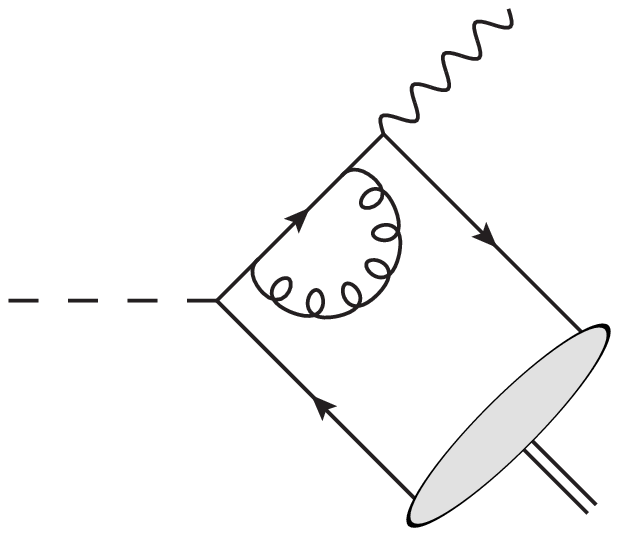}
\includegraphics[width=0.21\textwidth]{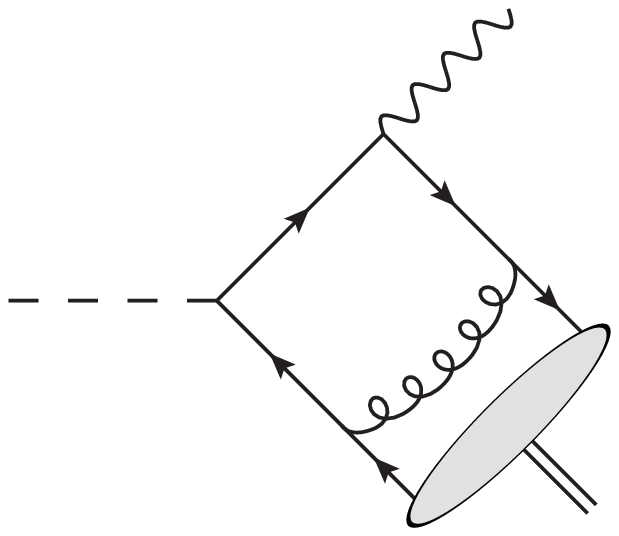}
\parbox{15.5cm}
{\caption{\label{fig:NLOgraphs}
One-loop QCD corrections to the first diagram in Figure~\ref{fig:diags}. Analogous corrections exist for the second diagram.}}
\end{center}
\end{figure}

The two leading-order graphs shown in Figure~\ref{fig:diags} are supplemented by the diagrams in Figure~\ref{fig:NLOgraphs}, which arise at ${\cal O}(\alpha_s)$. Including these effects is crucial in order to control the scale dependence of the transverse decay constant $f_V^\perp$, the Yukawa coupling $y_q$ and the LCDA $\phi_V^\perp$. It will also allow us to resum large logarithms of the form $\big(\alpha_s\ln(m_h^2/\mu_0^2)\big)^n$ to all orders in perturbation theory. Here $\mu_0\approx 1$\,GeV is a typical hadronic scale, at which model predictions for $f_V^\perp$ and $\phi_V^\perp$ are obtained. We work in dimensional regularization and subtract UV and IR divergences in the $\overline{\rm MS}$ scheme. The product of the bare decay constant times the LCDA of a transversely polarized vector meson is related to the product of the corresponding renormalized quantities via
\begin{equation}
   f_V^{\perp{\rm bare}}\,\phi_V^{\perp{\rm bare}}(x) 
   = \int_0^1\!dy\,Z_\phi^{-1}(x,y,\mu)\,f_V^\perp(\mu)\,\phi_V^\perp(y,\mu) \,,
\end{equation}
where at one-loop order
\begin{equation}
   Z_\phi(x,y,\mu) = \delta(x-y) + \frac{C_F\alpha_s(\mu)}{2\pi\epsilon}\,V_0^\perp(x,y) 
    + {\cal O}(\alpha_s^2) \,,
\end{equation}
with $C_F=4/3$. The relevant one-loop Brodsky-Lepage kernel reads \cite{Lepage:1979zb,Chernyak:1983ej} 
\begin{equation}\label{BLkernel}
   V_0^\perp(x,y) 
   = \frac12\,\delta(x-y) - \frac{1}{y(1-y)} \left[ x(1-y)\,\frac{\theta(y-x)}{y-x}
    + y(1-x)\,\frac{\theta(x-y)}{x-y} \right]_+ .
\end{equation}
For the decays $h\to V\gamma$, which are mediated by (pseudo-)scalar currents, an overall UV divergence remains, which is cancelled by the counterterm for the Yukawa coupling, derived from $y_{q,\rm bare}=\mu^\epsilon Z_y(\mu)\,y_q(\mu)$ with
\begin{equation}
   Z_y(\mu) = 1 - \frac{3C_F\alpha_s(\mu)}{4\pi\epsilon} + {\cal O}(\alpha_s^2) \,.
\end{equation}
When dealing with pseudo-scalar currents we employ the 't\,Hooft-Veltman (HV) scheme \cite{'tHooft:1972fi}, in which $\gamma_5=i\gamma^0\gamma^1\gamma^2\gamma^3$ anti-commutes with the four matrices $\gamma^\mu$ with $\mu\in\{0,1,2,3\}$, while it commutes with the remaining $(d-4)$ Dirac matrices $\gamma_\perp^\mu$. While this definition is mathematically consistent, it violates the Ward identities of chiral gauge theories by finite terms, which must be restored order by order in perturbation theory \cite{Bonneau:1980ya}. This is accomplished by performing the finite renormalization $P=Z_{\rm HV}^P P_{\rm HV}$ of the pseudo-scalar current $P=\bar q\gamma_5 q$, where \cite{Larin:1993tq}
\begin{equation}\label{ZHV}
   Z_{\rm HV}^P(\mu) = 1 - 2 C_F\,\frac{\alpha_s(\mu)}{\pi} + {\cal O}(\alpha_s^2) \,.
\end{equation}

By evaluating the relevant Feynman graphs in Figures~\ref{fig:diags} and \ref{fig:NLOgraphs}, we find that the direct contributions to the form factors in the amplitude decomposition (\ref{Ampl}) are given by
\begin{equation}\label{F1F2rela}
   F_{1,\rm direct}^V = \bar\kappa_V\,Q_V F_V \,, \qquad
   F_{2,\rm direct}^V = i\bar{\tilde\kappa}_V\,Q_V F_V \,, 
\end{equation}
where we have defined
\begin{equation}\label{kappaVdef}
   \bar\kappa_V 
   = \frac{1}{Q_V}\,\sum_q\,\bar\kappa_q\,Q_q\,\frac{f_V^{q\perp}}{f_V^\perp} \,, \qquad
   \bar{\tilde\kappa}_V 
   = \frac{1}{Q_V}\,\sum_q\,\bar{\tilde\kappa}_q\,Q_q\,\frac{f_V^{q\perp}}{f_V^\perp} \,.
\end{equation}
The reduced form factors $F_V$ are given by
\begin{equation}\label{FVres}
   F_V = \frac{m_b(\mu)}{v}\,\frac{f_V^\perp(\mu)}{f_V} 
    \int_0^1\!dx\,\frac{\phi_V^\perp(x,\mu)}{x(1-x)} 
    \left[ 1 + \frac{C_F\alpha_s(\mu)}{4\pi}\,h(x,m_h,\mu) + {\cal O}(\alpha_s^2) \right] ,
\end{equation}
with
\begin{equation}
   h(x,m_h,\mu) = 2\ln\big[x(1-x)\big] \left( \ln\frac{m_h^2}{\mu^2} - i\pi \right) 
    + \ln^2 x + \ln^2(1-x) - 3 \,.
\end{equation}
This result agrees with a previous calculation performed in \cite{Wang:2013ywc} apart from a typo.\footnote{These authors use the pole mass instead of the running quark mass in the prefactor, which adds $-3\ln(\mu^2/m_b^2)-4$ to the kernel $h(x,m_h,\mu)$. In eq.~(130) of \cite{Wang:2013ywc} one finds instead $-3\ln[\mu^2/(-m_h^2)]-4$. We are grateful to the authors for confirming this mistake.} 
We have expressed the Yukawa coupling in terms of the running $b$-quark mass using the second relation in (\ref{kappaq}). We focus primarily on the cases $V=J/\psi$ and $\Upsilon(nS)$, where to an excellent approximation the vector meson contains a single quark flavor $q$, and hence $\bar\kappa_{J/\psi}\approx\bar\kappa_c$ and $\bar\kappa_{\Upsilon(nS)}\approx\kappa_b$, and similarly for the CP-odd parameters $\bar{\tilde\kappa}_V$. For the light mesons $V=\rho^0$, $\omega$ and $\phi$, on the other hand, flavor-mixing effects can be important. This concerns, in particular, the possibility of a small admixture of an $|s\bar s\rangle$ flavor component in the wave functions of $\rho^0$ and $\omega$, which can be important due to the smallness of the Yukawa couplings to the up and down quarks in the SM. Since the $\rho^0$ meson is a member of an isospin triplet, its flavor mixing into $|s\bar s\rangle$ can only be caused by electromagnetic interactions or isospin-violating effects in QCD. Both types of effects are estimated to be very small, and hence we expect that $|f_{\rho^0}^{s\perp}/f_{\rho^0}^\perp|\lll 1$. To good approximation we can thus use the naive relation  
\begin{equation}\label{effectivecharges}
   \bar\kappa_{\rho^0}\approx \frac{2\bar\kappa_u+\bar\kappa_d}{3} 
    ~~ \stackrel{\rm SM}{\to} ~~ 6.1\cdot 10^{-4} \,.
\end{equation}
The situation is different for the case of the $\omega$ meson, whose mixing into an $|s\bar s\rangle$ flavor state can be non-negligible. In Appendix~\ref{app:fVeff} we derive explicit expressions for the parameters $\bar\kappa_\omega$ and $\bar\kappa_\phi$ in a simple flavor-mixing scheme for the $\omega\!-\!\phi$ system. Assuming that $|\bar\kappa_s|\gg|\bar\kappa_{u,d}|$ like in the SM, and working in the SU(3) limit and to first order in the small mixing angle $\theta_{\omega\phi}$, we obtain
\begin{equation}\label{rhoomegamixed}
\begin{aligned}
   \bar\kappa_\omega 
   &\approx 2\bar\kappa_u-\bar\kappa_d + \sqrt2\,\bar\kappa_s\,\theta_{\omega\phi}(m_\omega^2) 
   &&\stackrel{\rm SM}{\to} ~~
    \big( - 0.08 + 26.8\,\theta_{\omega\phi} \big) \cdot 10^{-3} \,, \\
   \bar\kappa_\phi
   &\approx \bar\kappa_s \left[ 1 + \frac{\theta_{\omega\phi}(m_\phi^2)}{\sqrt2} \right] 
   &&\stackrel{\rm SM}{\to} ~~~
    0.019 + 0.013\,\theta_{\omega\phi} \,.
\end{aligned}
\end{equation}
In the SM the contributions from the up and down quarks almost precisely cancel in $\bar\kappa_\omega$, and hence the contribution induced by $\omega\!-\!\phi$ mixing is likely to be the dominant one. Existing estimates for the mixing angle $\theta_{\omega\phi}$ derived from mass-independent analyses include $\theta_{\omega\phi}\approx 0.05$ \cite{Shifman:1978by} and $\theta_{\omega\phi}\approx 0.06$ \cite{Benayoun:1999fv,Kucukarslan:2006wk}. On the other hand, in a more recent mass-dependent analysis the values $\theta_{\omega\phi}(m_\omega^2)\approx 0.008$ and $\theta_{\omega\phi}(m_\phi^2)\approx 0.081$ were obtained \cite{Benayoun:2007cu}. We conclude from this discussion that $\bar\kappa_\phi\approx\bar\kappa_s$ to good approximation, while a more accurate description of flavor-mixing effects would be required before the quantity $\bar\kappa_\omega$ can be interpreted reliably in terms of quark Yukawa couplings.

In the factorization formula (\ref{FVres}) all non-perturbative hadronic physics is contained in the decay constants and the LCDA. The quantity multiplying the LCDA under the integral is the hard-scattering coefficient, which can be calculated in perturbation theory. It depends on the momentum distribution of the quark inside the hadronic bound state. QCD-based model calculations of LCDAs are typically performed at a low hadronic scale $\mu_0\sim 1$\,GeV. If such a low value is chosen for the factorization scale $\mu$ in (\ref{FVres}), the hard-scattering coefficient contains large logarithms of the form $\big(\alpha_s\ln(m_h^2/\mu_0^2)\big)^n$ with $\ln(m_h^2/\mu_0^2)\approx 9.7$, which should be resummed to all orders in perturbation theory. To perform this resummation, it is convenient to use the expansion of the LCDA in the basis of Gegenbauer polynomials, which reads \cite{Lepage:1979zb,Chernyak:1983ej}
\begin{equation}\label{Gegenbauer}
   \phi_V^\perp(x,\mu) 
   = 6x(1-x)\,\bigg[ 1 + \sum_{n=1}^\infty a_n^{V_\perp}(\mu)\,C_n^{(3/2)}(2x-1) \bigg] \,.
\end{equation}
We can then rewrite (\ref{FVres}) as a sum over Gegenbauer moments, finding 
\begin{equation}\label{FVres2}
   F_V = \frac{6m_b(\mu)}{v}\,\frac{f_V^\perp(\mu)}{f_V}
    \left[ 1 - \frac{C_F\alpha_s(\mu)}{\pi}\,\ln\frac{m_h^2}{\mu^2} \right] I_V(m_h) \,,
\end{equation}
with
\begin{equation}\label{IVres}
   I_V(m_h) = \sum_{n=0}^\infty C_{2n}(m_h,\mu)\,a_{2n}^{V_\perp}(\mu) \,.
\end{equation}
The factor in brackets in (\ref{FVres2}) precisely compensates the scale dependence of the product $m_b(\mu)\,f_V^\perp(\mu)$, while the quantity $I^V$ is formally scale invariant. Using a technique developed in \cite{Grossman:2015lea}, we obtain for the hard-scattering coefficients $C_n$ in moment space the closed expression
\begin{equation}
   C_n(m_h,\mu) = 1 + \frac{C_F\alpha_s(\mu)}{4\pi}
    \left[ -4 \left( H_{n+1} - 1 \right) \left( \ln\frac{m_h^2}{\mu^2} - i\pi \right)
    + 4 H_{n+1}^2 - 3 + 4i\pi \right] + {\cal O}(\alpha_s^2) \,,
\end{equation}
where $H_{n+1}=\sum_{k=1}^{n+1}\frac{1}{k}$ are the harmonic numbers. As a consequence of the symmetry of the hard-scattering coefficient under the exchange $x\leftrightarrow(1-x)$, the sum in (\ref{IVres}) runs over even Gegenbauer moments only. Large logarithms of the type $\big(\alpha_s\ln\frac{m_h^2}{\mu^2}\big)^n$ can now be resummed readily by choosing the factorization scale of order $\mu\sim m_h$ and evolving the scale-dependent quantities $m_b(\mu)$, $f_V^\perp(\mu)$ and $a_n^{V_\perp}(\mu)$ up to that scale. The solution of the corresponding renormalization-group (RG) equations at NLO of RG-improved perturbation theory is discussed in Appendix~\ref{app:RGEs}. At leading order in QCD the Gegenbauer moments $a_n^{V_\perp}(\mu)$ in (\ref{Gegenbauer}) are renormalized multiplicatively, such that \cite{Lepage:1979zb,Shifman:1980dk}
\begin{equation}\label{anevol}
   a_n^{V_\perp}(\mu) 
   = \left( \frac{\alpha_s(\mu)}{\alpha_s(\mu_0)} \right)^{\frac{\gamma_n^\perp}{2\beta_0}} 
    a_n^{V_\perp}(\mu_0) \,, 
   \qquad \mbox{with} \quad
   \gamma_n^\perp = 8 C_F \left( H_{n+1} - 1 \right) .
\end{equation}
Here $\mu_0$ denotes a low reference scale, while $\mu={\cal O}(m_h)$ is the hard factorization scale, to which the LCDAs are evolved. The NLO corrections to these relations are discussed in Appendix~\ref{app:RGEs}. They have a negligible impact on our numerical results. All of the anomalous dimensions are strictly positive (for $n\ne 0$), which implies that $a_n^{V_\perp}(\mu)\to 0$ in the formal limit $\mu\to\infty$. In this limit the leading-twist LCDAs approach the asymptotic form $6x(1-x)$. Similarly, it follows from relation (\ref{RGEf}) in Appendix~\ref{app:RGEs} that the transverse decay constants of vector mesons vanish in the asymptotic limit, i.e.\ $f_V^\perp(\mu)\to 0$ for $\mu\to\infty$.

\begin{table}
\begin{center}
\begin{tabular}{|c|ccc|cc|}
\hline 
Meson $V$ & $f_V$~[MeV] & $f_V^\perp(2\,{\rm GeV})/f_V$ & $a_2^{V_\perp}(\mu_0)$ 
 & $Q_V$ & $v_V$ \\
\hline 
$\rho^0$ & $216.3\pm 1.3$ & $0.72\pm 0.04$ &  $0.14\pm 0.06$ 
 & $\frac{1}{\sqrt2}$ & $\frac{1}{\sqrt2}\left( \frac12 - s_W^2 \right)$ \\  
$\omega$ & $194.2\pm 2.1$ & $0.71\pm 0.05$ & $0.15\pm 0.07$ 
 & $\frac{1}{3\sqrt2}$ & $-\frac{s_W^2}{3\sqrt2}$ \\
$\phi$ & $223.0\pm 1.4$ & $0.76\pm 0.04$ & $0.14\pm 0.07$ 
 & $-\frac13$ & $-\frac14+\frac{s_W^2}{3}$ \\
\hline 
\end{tabular}
\parbox{15.5cm}
{\caption{\label{tab:hadronic_inputs} 
Hadronic input parameters for light vector mesons. The decay constants $f_V$ are extracted from data on the electromagnetic decay widths $V\to l^+ l^-$ \cite{Grossman:2015lea}, while the ratios $f_V^\perp/f_V$ are derived from a compilation of theoretical predictions. The values of the Gegenbauer moments at the scale $\mu_0=1$\,GeV are taken from \cite{Ball:2007rt,Ball:2006eu}. The last two columns show the effective charges $Q_V$ and $v_V$ defined in (\ref{fVdef}) and below (\ref{F1Vindir}).}}
\end{center}
\end{table} 

It has been emphasized in \cite{Grossman:2015lea} that RG evolution effects render our predictions rather insensitive to the precise values of the Gegenbauer moments. From (\ref{IVres}), we obtain
\begin{equation}\label{IVnumerics}
\begin{aligned}
   \mbox{Re}\,I_V(m_h) &= 1.01 + 1.13 a_2^{V_\perp}(m_h) + 1.21 a_4^{V_\perp}(m_h)
    + 1.29 a_6^{V_\perp}(m_h) + 1.35 a_8^{V_\perp}(m_h) + \dots \\
   &\approx 1.01 + 0.51 a_2^{V_\perp}(\mu_0) + 0.36 a_4^{V_\perp}(\mu_0)
    + 0.29 a_6^{V_\perp}(\mu_0) + 0.24 a_8^{V_\perp}(\mu_0) + \dots \,.
\end{aligned}
\end{equation}
While all Gegenbauer moments have ${\cal O}(1)$ coefficients at the high-energy scale, the coefficients of the higher moments are strongly reduced when one expresses the answer in terms of moments normalized at the low scale $\mu_0=1$\,GeV.

In order to obtain numerical predictions for the reduced form factors we need as hadronic input parameters the decay constants $f_V$ and $f_V^\perp$ and the Gegenbauer moments $a_{2n}^{V_\perp}$ of the various vector mesons. As mentioned earlier, the decay constants $f_V$ can be extracted from experimental data, and up-to-date values have been derived in \cite{Grossman:2015lea}. The ratios $f_V^\perp/f_V$ needed in (\ref{FVres2}) must be obtained using some non-perturbative approach, such as lattice QCD, light-cone QCD sum rules or the non-relativistic effective theory NRQCD for heavy quarkonia \cite{Caswell:1985ui,Bodwin:1994jh}, which provides a systematic expansion of hadronic matrix elements in powers of the small velocity $v\sim\alpha_s(m_Q v)$ of the heavy quark in the quarkonium rest frame. Details of such determinations are reviewed in Appendix~\ref{app:fVperp}. In Table~\ref{tab:hadronic_inputs} we compile the relevant input parameters for light vector mesons. Because of the lack of information about higher Gegenbauer moments we can only keep few terms in the infinite sum (\ref{Gegenbauer}). The systematics of the Gegenbauer expansion has been discussed in \cite{Grossman:2015lea}, where it was pointed out that the higher moments $a_n^{V_\perp}$ with $n\gg 1$ fall off faster than $1/n$. Indeed, high-rank Gegenbauer polynomials $C_n^{(3/2)}(2x-1)$ with $n\gg 1$ would resolve structures on scales $\Delta x\sim 1/n$. For a light vector meson $V$, it is reasonable to assume that the LCDA $\phi_V^\perp(x)$ does not exhibit pronounced structures at scales much smaller than ${\cal O}(1)$. To estimate the impact of higher moments we use $a_4^{V_\perp}(\mu_0)=\pm 0.15$ for our error estimates. Relation (\ref{IVnumerics}) suggests that the effect of yet higher-order terms is small. 

\begin{table}
\begin{center}
\begin{tabular}{|c|ccc|cc|}
\hline 
Meson $V$ & $f_V$~[MeV] & $f_V^\perp(2\,{\rm GeV})/f_V$ & $\sigma_V(\mu_0)$ 
 & $Q_V$ & $v_V$ \\
\hline 
$J/\psi$ & $403.3\pm 5.1$ & $0.91\pm 0.14$ & $0.228\pm 0.005\pm 0.057$ 
 & $\frac23$ & $\frac14-\frac{2s_W^2}{3}$ \\ 
$\Upsilon(1S)$ & $684.4\pm 4.6$ & $1.09\pm 0.04$ & $0.112\pm 0.004\pm 0.028$ 
 & $-\frac13$ & $-\frac14+\frac{s_W^2}{3}$ \\ 
$\Upsilon(2S)$ & $475.8\pm 4.3$ & $1.08\pm 0.05$ & $0.144\pm 0.007\pm 0.036$ 
 & $-\frac13$ & $-\frac14+\frac{s_W^2}{3}$ \\ 
$\Upsilon(3S)$ & $411.3\pm 3.7$ & $1.07\pm 0.05$ & $0.162\pm 0.010\pm 0.041$ 
 & $-\frac13$ & $-\frac14+\frac{s_W^2}{3}$ \\ 
\hline 
\end{tabular}
\parbox{15.5cm}
{\caption{\label{tab:hadronic_inputs2} 
Hadronic input parameters for heavy quarkonium states. The decay constants $f_V$ are extracted from data on the electromagnetic decay widths $V\to l^+ l^-$ \cite{Grossman:2015lea}, while the ratios $f_V^\perp/f_V$ are derived from NRQCD scaling relations. The width parameters $\sigma_V$ are obtained from relation (\ref{mom2}), where the first error is of parametric origin and the second one parameterizes the uncertainty due to higher-order effects. The last two columns show the effective charges $Q_V$ and $v_V$.}}
\end{center}
\end{table} 

The LCDAs of heavy mesons exhibit a different behavior, since the presence of the heavy-quark mass introduces a new scale. For a quarkonium state $V\sim (Q\bar Q)$ composed of two identical heavy quarks, the LCDA peaks at $x=1/2$ and has a width that tends to zero in the limit of infinite heavy-quark mass. The second moment of the LCDA around $x=1/2$ can be related to a local NRQCD matrix element called $\langle v^2\rangle_V$ \cite{Braguta:2006wr}. Including the one-loop QCD corrections calculated in \cite{Wang:2013ywc}, we obtain
\begin{equation}\label{mom2}
   4\sigma_V^2(\mu) \equiv 
   \int_0^1\!dx \left( 2x - 1 \right)^2 \phi_V^\perp(x,\mu)
   = \frac{\langle v^2\rangle_V}{3} 
    + \frac{C_F\alpha_s(\mu)}{4\pi} \left( \frac{28}{9} - \frac23 \ln\frac{m_Q^2}{\mu^2} \right) 
    + \dots \,.
\end{equation}
A critical discussion of the extraction of the parameters $\langle v^2\rangle_V$ for different quarkonium states is presented in Appendix~\ref{app:fVperp}. Using the values compiled there, but with increased error estimates, we obtain the ratios of decay constants and the width parameters $\sigma_V(\mu_0)$ at the low scale $\mu_0=1$\,GeV shown in Table~\ref{tab:hadronic_inputs2}. As a reasonable model at the scale $\mu_0$ we adopt the form \cite{Grossman:2015lea}
\begin{equation}\label{LCDAQQ}
   \phi_V^\perp(x,\mu_0) = N_\sigma\,\frac{4x(1-x)}{\sqrt{2\pi}\sigma_V}\,
    \exp\left[ - \frac{(x-\frac12)^2}{2\sigma_V^2} \right] ,
\end{equation}
where the polynomial in front of the Gaussian factor ensures that the LCDA vanishes at the endpoints $x=0,1$. In order to estimate the uncertainties related to the functional form and to capture the effects of unknown higher-order corrections to relation (\ref{mom2}), we include a second error of $\pm 25\%$ on the $\sigma_V$ parameters. Given this form, we compute the first 20 Gegenbauer moments at the low scale $\mu_0$, evolve them up to the factorization scale $\mu\approx m_h$ using (\ref{anevol}), and use these results in evaluating relation (\ref{IVres}). 

\begin{table}
\begin{center}
\begin{tabular}{|c|c|c|}
\hline 
Meson & Form factor with errors [\%] & Combined value [\%] \\ 
\hline 
$F_{\rho^0}$ & $4.30\,_{-\,0.05}^{+\,0.04}\,{}_\mu\pm 0.03_{m_b}\pm 0.24_f\pm 0.12_{a_2}\pm 0.22_{a_4}$ 
 & $(4.30\pm 0.35) + i(0.67\pm 0.14)$ \\ 
 & $+i\big(0.67\,_{-\,0.10}^{+\,0.14}\,{}_\mu\pm 0.00_{m_b}\pm 0.04_f\pm 0.03_{a_2}\pm 0.06_{a_4}\big)$ 
 & \\ 
\hline 
$F_\omega$ & $4.26\,_{-\,0.05}^{+\,0.04}\,{}_\mu\pm 0.03_{m_b}\pm 0.30_f\pm 0.14_{a_2}\pm 0.21_{a_4}$ 
 & $(4.26\pm 0.40) + i(0.66\pm 0.14)$ \\ 
 & $+i\big(0.66\,_{-\,0.10}^{+\,0.14}\,{}_\mu\pm 0.00_{m_b}\pm 0.05_f\pm 0.03_{a_2}\pm 0.06_{a_4}\big)$ 
 & \\ 
\hline
$F_\phi$ & $4.53\,_{-\,0.05}^{+\,0.04}\,{}_\mu\pm 0.03_{m_b}\pm 0.24_f\pm 0.15_{a_2}\pm 0.23_{a_4}$ 
 & $(4.53\pm 0.37) + i(0.70\pm 0.15)$ \\ 
 & $+i\big(0.70\,_{-\,0.10}^{+\,0.14}\,{}_\mu\pm 0.01_{m_b}\pm 0.04_f\pm 0.04_{a_2}\pm 0.06_{a_4}\big)$ 
 & \\ 
\hline
$F_{J/\psi}$ & $4.54\,_{-\,0.04}^{+\,0.02}\,{}_\mu\pm 0.03_{m_b}\pm 0.70_f\,_{-\,0.17}^{+\,0.13}\,{}_{\sigma_V}$
 & $(4.54\pm 0.72) + i(0.63\pm 0.14)$ \\ 
 & $+i\big(0.63\,_{-\,0.08}^{+\,0.11}\,{}_\mu\pm 0.00_{m_b}\pm 0.10_f\,_{-\,0.04}^{+\,0.03}\,{}_{\sigma_V}\big)$ 
 & \\ 
\hline
$F_{\Upsilon(1S)}$ & $5.04\,_{-\,0.03}^{+\,0.02}\,{}_\mu\pm 0.04_{m_b}\pm 0.18_f\,_{-\,0.07}^{+\,0.09}\,{}_{\sigma_V}$
 & $(5.04\pm 0.21) + i(0.66\pm 0.10)$ \\ 
 & $+i\big(0.66\,_{-\,0.08}^{+\,0.12}\,{}_\mu\pm 0.00_{m_b}\pm 0.02_f\,_{-\,0.01}^{+\,0.02}\,{}_{\sigma_V}\big)$ 
 & \\ 
\hline 
$F_{\Upsilon(2S)}$ & $5.09\,_{-\,0.04}^{+\,0.02}\,{}_\mu\pm 0.04_{m_b}\pm 0.24_f\,_{-\,0.12}^{+\,0.13}\,{}_{\sigma_V}$
 & $(5.09\pm 0.27) + i(0.68\pm 0.11)$ \\ 
 & $+i\big(0.68\,_{-\,0.09}^{+\,0.12}\,{}_\mu\pm 0.00_{m_b}\pm 0.03_f\,_{-\,0.02}^{+\,0.03}\,{}_{\sigma_V}\big)$ 
 & \\ 
\hline 
$F_{\Upsilon(3S)}$ & $5.11\,_{-\,0.04}^{+\,0.02}\,{}_\mu\pm 0.04_{m_b}\pm 0.24_f\,_{-\,0.14}^{+\,0.15}\,{}_{\sigma_V}$
 & $(5.11\pm 0.29) + i(0.69\pm 0.12)$ \\ 
 & $+i\big(0.69\,_{-\,0.09}^{+\,0.12}\,{}_\mu\pm 0.00_{m_b}\pm 0.03_f\,_{-\,0.03}^{+\,0.04}\,{}_{\sigma_V}\big)$ 
 & \\ 
\hline 
\end{tabular}
\parbox{15.5cm}
{\caption{\label{tab:FVnumbers} 
Theory predictions for the reduced form factors $F_V$ including error estimates.}}
\end{center}
\end{table} 

We are now ready to present our numerical results for the direct contributions to the reduced form factors $F_V$ in (\ref{FVres2}) for various vector mesons, including detailed error estimates. They are collected in Table~\ref{tab:FVnumbers}. The different sources of theoretical errors contain a perturbative uncertainty (subscript ``$\mu$''), which we determine by varying the factorization scale $\mu$ between $m_h/2$ and $2m_h$. Once the NLO corrections are included our results are very stable under scale variations. The scale uncertainties are larger for the imaginary parts than for the real parts of the form factors, since these start at ${\cal O}(\alpha_s)$ and there is thus no compensation of the scale dependence. We emphasize, however, that the imaginary parts only have a small impact on our numerical predictions for the decay rates. We also include the uncertainty in the value of the $b$-quark mass, which has a very small impact. The uncertainties related to hadronic parameters include the ratio $f_V^\perp/f_V$ (subscript ``$f$'') and uncertainties in the shapes of the LCDAs, as modeled by the values of the Gegenbauer moments $a_2^{V_\perp}$ and $a_4^{V_\perp}$ (light mesons) and the width parameter $\sigma_V$ (heavy mesons). These hadronic uncertainties are the dominant sources of errors. The last column in the table shows the results obtained when all errors are added in quadrature. These numbers will be used for our phenomenological analysis in Section~\ref{sec:pheno}. We observe that the spread of the results for the form factors $F_V$ for different vector mesons is rather small. The theoretical uncertainties on the real part of $F_V$ are typically between 4\% and 9\%. The only exception is $F_{J/\psi}$, for which the uncertainty in the ratio of decay constants is about 16\%. It would probably be possible to reduce this uncertainty by performing a more detailed NRQCD analysis.

\begin{figure}
\begin{center}
\begin{tabular}{rcl}
\includegraphics[width=0.29\textwidth]{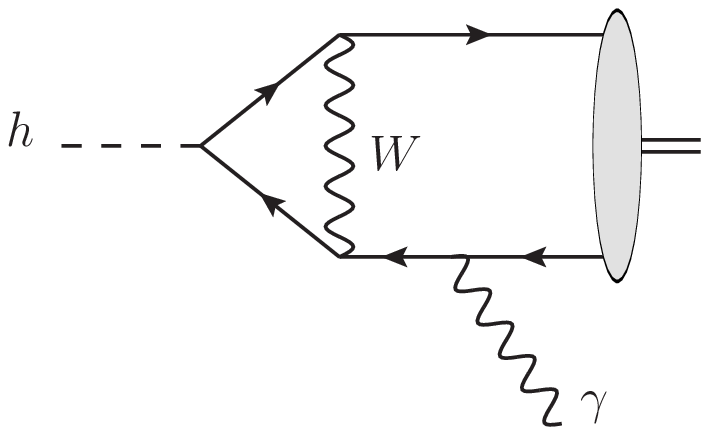} &&
\includegraphics[width=0.29\textwidth]{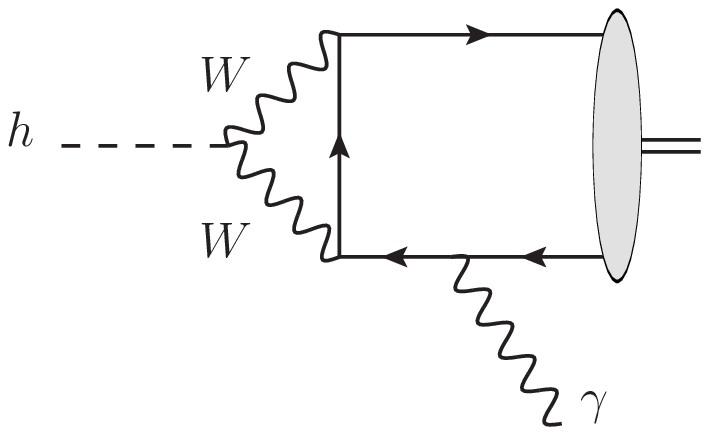} \\[2mm]
\includegraphics[width=0.23\textwidth]{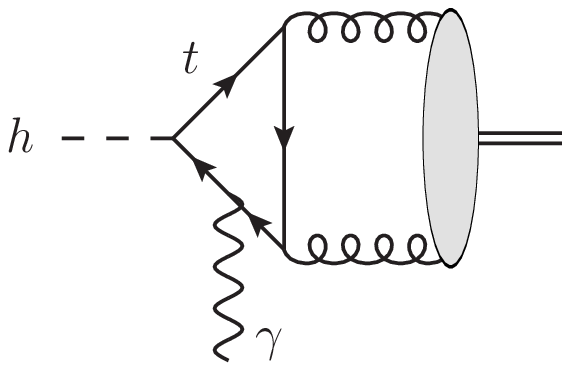} &&
\includegraphics[width=0.32\textwidth]{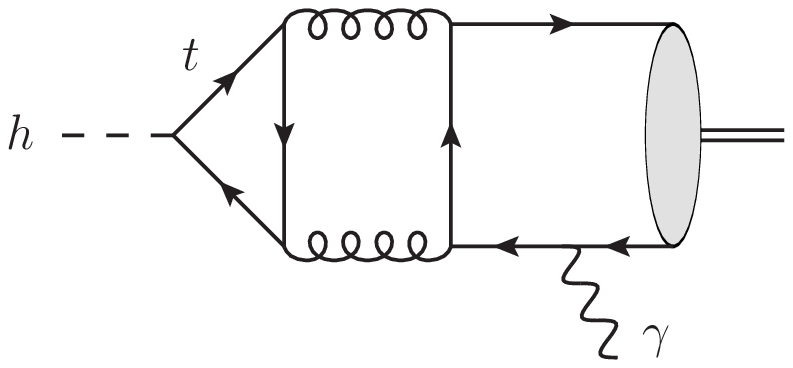}
\end{tabular}
\parbox{15.5cm}
{\caption{\label{fig:EWcors}
Examples of electroweak radiative corrections (top row) and higher-order QCD radiative corrections (bottom row) to the $h\to V\gamma$ decay amplitudes.}}
\end{center}
\end{figure}

Power-suppressed corrections to our results (\ref{FVres}) and (\ref{FVres2}) can be organized in an expansion in $(\Lambda_{\rm QCD}/m_h)^2$ for light mesons and $(m_V/m_h)^2$ for heavy mesons \cite{Grossman:2015lea}. They are at most of order $10^{-2}$ for $V=\Upsilon(nS)$, $10^{-3}$ for $V=J/\psi$, and $10^{-4}$ for all light mesons. It is thus safe to work with the leading-order terms. In our analysis we neglect two-loop QCD corrections, whose effects should be covered by the error we estimate from scale variations, and one-loop QED or electroweak radiative corrections, a few examples of which are shown in the top row in Figure~\ref{fig:EWcors}. For flavor-diagonal final-state mesons, the first diagram involves a factor $\frac{m_W}{v}\,\frac{\alpha}{\pi}\sim 0.7\cdot 10^{-3}$ instead of the Yukawa coupling $y_q$ in the diagrams in Figure~\ref{fig:diags}, while the second diagram involves a factor $y_t\,\frac{\alpha}{\pi}\sim 2\cdot 10^{-3}$. This is smaller (by roughly a factor 10) than the charm-quark Yukawa coupling and of the same order as the strange-quark Yukawa coupling. If the goal is to reach sensitivity to the strange-quark Yukawa coupling in the SM, then these electroweak corrections should be calculated. However, such a level of sensitivity will be out of reach at the LHC. In the bottom row in Figure~\ref{fig:EWcors} we show other examples of neglected diagrams, which involve a $h\to gg(\gamma)$ transition followed by the conversion of the two gluons into the final-state meson. The diagram on the left corresponds to a two-gluon LCDA of the vector meson, which does not exist at leading twist due to the Landau-Yang theorem. The graph on the right is analogous to the second diagram shown in the first row, but with the internal $W$ bosons replaced by gluons. A naive estimate indicates that the two types of effects should be of similar magnitude. 

\subsection{Indirect contributions to the form factors}
\label{subsec:indirect}

\begin{figure}
\begin{center}
\includegraphics[width=0.23\textwidth]{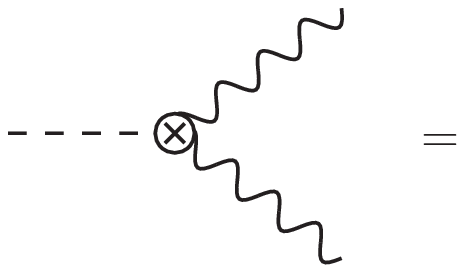} \hspace{-4mm}
\includegraphics[width=0.19\textwidth]{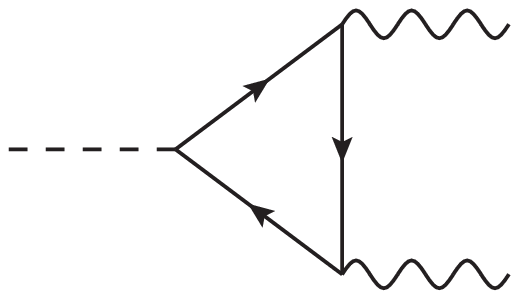} \quad
\includegraphics[width=0.19\textwidth]{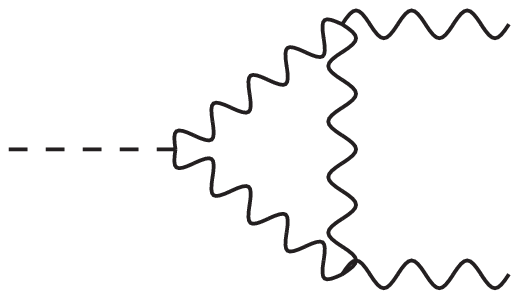}\quad
\includegraphics[width=0.19\textwidth]{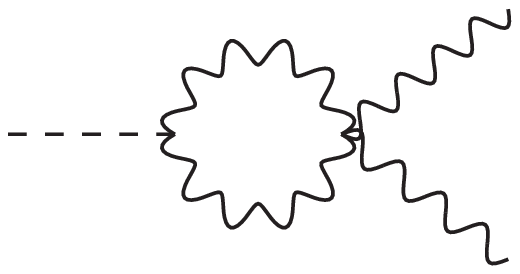}
\parbox{15.5cm}
{\caption{\label{fig:Leffhgaga}
One-loop SM contributions to the effective $h\gamma\gamma$ and $h\gamma Z$ vertices.}}
\end{center}
\end{figure}

We now proceed to study the photon- and $Z$-pole contributions to the decay amplitude shown in the third diagram in Figure~\ref{fig:diags}. The crossed circle in this diagram represents the off-shell $h\to\gamma\gamma^*$ and $h\to\gamma Z^*$ amplitudes, which in the SM are induced by loop graphs involving a virtual charged fermion or a $W$ boson (in unitary gauge), as shown in Figure~\ref{fig:Leffhgaga}. Since the indirect contributions to the decay amplitudes are numerically dominant over the direct ones and our goal is to compute them with the highest possible accuracy, we include the effect that the intermediate gauge boson is slightly off shell ($k^2=m_V^2$), and we keep the full dependence on the meson mass even though this is a very small effect. 

The exact one-loop expressions for the off-shell $h\to\gamma\gamma^*$ and $h\to\gamma Z^*$ amplitudes have been derived in \cite{Bergstrom:1985hp}. Using these results and extending them to the case of the pseudo-scalar Higgs couplings in (\ref{eqn:effL}), we find
\begin{equation}\label{F1Vindir}
\begin{aligned}
   F_{1,\rm indirect}^V 
   &= \phantom{i\,} \frac{\alpha(m_V)}{\pi}\,\frac{m_h^2-m_V^2}{m_V\,v} 
    \left[ Q_V\,C_{\gamma\gamma}(x_V) 
    - \frac{v_V}{\left(s_W c_W\right)^2}\,\frac{m_V^2}{m_Z^2-m_V^2}\,
    C_{\gamma Z}(x_V) \right] , \\
   F_{2,\rm indirect}^V 
   &= i\,\frac{\alpha(m_V)}{\pi}\,\frac{m_h^2-m_V^2}{m_V\,v} 
    \left[ Q_V\,\tilde C_{\gamma\gamma}(x_V) 
    - \frac{v_V}{\left(s_W c_W\right)^2}\,\frac{m_V^2}{m_Z^2-m_V^2}\,
    \tilde C_{\gamma Z}(x_V) \right] ,
\end{aligned}
\end{equation}
where $x_V=m_V^2/m_h^2$ accounts for the effects of the off-shell boson, and $v_V\equiv\sum_q c_q^V v_q$ is defined in analogy with $Q_V$ in (\ref{fVdef}), where $v_f=\frac12\,T_3^f-s_W^2 Q_f$ are the vector couplings of the $Z$ boson to fermions. It is a safe approximation to neglect flavor-mixing effects for the subleading contribution from $h\to\gamma Z^*\to\gamma V$. At one-loop order, the loop functions are given by
\begin{equation}\label{Cgaga}
\begin{aligned}
   C_{\gamma\gamma}(x_V) 
   &= \sum_q \kappa_q\,\frac{2 N_c Q_q^2}{3}\,A_f(\tau_q,x_V)
    + \sum_l \kappa_l\,\frac{2 Q_l^2}{3}\,A_f(\tau_l,x_V)
    - \frac{\kappa_W}{2}\,A_W^{\gamma\gamma}(\tau_W,x_V) + \kappa_{\gamma\gamma} \,, \\
   C_{\gamma Z}(x_V) 
   &= \sum_q \kappa_q\,\frac{2 N_c Q_q v_q}{3}\,A_f(\tau_q,x_V)
    + \sum_l \kappa_l\,\frac{2 Q_l v_l}{3}\,A_f(\tau_l,x_V)
    - \frac{\kappa_W}{2}\,A_W^{\gamma Z}(\tau_W,x_V) + \kappa_{\gamma Z} \,,
\end{aligned}    
\end{equation}
and
\begin{equation}\label{Cgagatilde}
\begin{aligned}
   \tilde C_{\gamma\gamma}(x_V) 
   &= \sum_q \tilde\kappa_q N_c Q_q^2\,B_f(\tau_q,x_V)
    + \sum_l \tilde\kappa_l\,Q_l^2\,B_f(\tau_l,x_V) + \tilde\kappa_{\gamma\gamma} \,, \\
   \tilde C_{\gamma Z}(x_V) 
   &= \sum_q \tilde\kappa_q N_c Q_q v_q\,B_f(\tau_q,x_V)
    + \sum_l \tilde\kappa_l\,Q_l v_l\,B_f(\tau_l,x_V) + \tilde \kappa_{\gamma Z} \,.
\end{aligned}    
\end{equation}
The first two terms in each coefficient are the contributions from the quarks and leptons, the third term in $C_{\gamma\gamma}$ and $C_{\gamma Z}$ arises from gauge-boson loops, and the last term accounts for possible new-physics contributions parameterized by the operators shown in the second line of (\ref{eqn:effL}). We have introduced the dimensionless variables $\tau_f=4m_f^2/m_h^2$ (for $f=q,l$) and $\tau_W=4m_W^2/m_h^2$. We use the running quark masses $m_q(m_h)$ when evaluating the variables $\tau_q$, which is appropriate in view of the large momentum transfer in the loop. Explicit expressions for the loop functions $A_f$, $A_W^{\gamma V}$ and $B_f$ are given in Appendix~\ref{app:loopfunctions}. In the SM we have $\kappa_q=\kappa_l=\kappa_W=1$ and $\kappa_{\gamma\gamma}=\kappa_{\gamma Z}=0$. The effective Higgs couplings $\tilde\kappa_i$ entering in (\ref{Cgagatilde}) all vanish in the SM. 

Since for small values of $\tau_f$ the fermion loop functions $A_f(\tau_f,x_V)$ and $B_f(\tau_f,x_V)$ are proportional to $\tau_f$, it suffices for all practical purposes to keep the contributions from the third-generation fermions. The effects of the off-shellness of the photon that converts into the final-state vector meson gives rise to very small corrections. The relevant variable $x_V=m_V^2/m_h^2$ varies between $3.8\times 10^{-5}$ for $V=\rho$ and $6.9\times 10^{-3}$ for $\Upsilon(3S)$. Note also that the contribution of the $h\to\gamma Z^*$ amplitude in (\ref{F1Vindir}) is by itself strongly power suppressed. Numerically, we obtain
\begin{equation}
\begin{aligned}
   C_{\gamma\gamma}(0) 
   &= \kappa_{\gamma\gamma} - 4.164 \kappa_W + 0.920 \kappa_t - (0.012 - 0.011 i) \kappa_\tau
    - (0.007 - 0.008 i) \kappa_b \\
   &\quad\mbox{}- (0.015 - 0.010 i) \bar\kappa_c - 0.001 \bar\kappa_s + \dots \,, \\
   C_{\gamma Z}(0) 
   &= \kappa_{\gamma Z} - 2.173 \kappa_W + 0.132 \kappa_t - (0.004 - 0.004 i) \kappa_b 
    - (0.002 - 0.001 i) \bar\kappa_c + \dots \,,
\end{aligned}
\end{equation}
and similar expressions hold for the CP-odd coefficients $\tilde C_{\gamma\gamma}$ and $\tilde C_{\gamma Z}$. Notice that the contributions from second-generation fermions are very small even if one assumes that their Yukawa couplings are as large as those of the $b$-quark (i.e., for $\bar\kappa_{c,s}=1$). In the SM, we have $C_{\gamma\gamma}(0)=-3.266+0.021i$ and $C_{\gamma Z}(0)=-2.046+0.005i$, while the CP-odd coefficients $\tilde C_{\gamma\gamma}$ and $\tilde C_{\gamma Z}$ vanish. In the expressions for the form factors in (\ref{F1Vindir}) these coefficients are weighted by different factors. Note also that to good approximation the indirect contributions to the form factors are proportional to $1/m_V$ and hence they are larger for lighter mesons. 

Let us now discuss the impact of QCD and electroweak radiative corrections on the above results. Gluon exchanges between the two quark lines that make up the final-state meson in the third diagram in Figure~\ref{fig:diags} are of non-perturbative nature and are accounted for by the meson decay constant $f_V$. Radiative corrections connecting the effective $h\gamma\gamma$ and $h\gamma Z$ vertices in the last diagram in Figure~\ref{fig:diags} to the final-state quarks give rise to graphs which no longer receive the enhancement proportional to $m_h^2/m_V^2$ from the photon propagator. Also, at least two gluons would need to be exchanged by color conservation. The effect of such diagrams is negligible. We thus only need to worry about QCD corrections to the $h\to\gamma\gamma^*$ and $h\to\gamma Z^*$ amplitudes, which arise from two-loop graphs in which a gluon is exchanged inside the quark loop. These corrections have been calculated in numerical form in \cite{Djouadi:1990aj} and analytically in \cite{Spira:1995rr}. In practice the corrections are only relevant for the top-quark contribution. Their effect is to enhance the decay amplitude by a few percent. Two-loop electroweak corrections to the $h\to\gamma\gamma$ amplitude in the SM were calculated in \cite{Aglietti:2004nj,Degrassi:2005mc} (see also \cite{Actis:2008ts}). One finds that they are small and negative. For $m_h=125.09$\,GeV, the effects of QCD and electroweak corrections nearly cancel each other, leaving a total correction of about $-0.2\%$ \cite{Degrassi:2005mc}. In our phenomenological analysis we will ignore radiative corrections on the indirect contributions computed here, as their combined effect is well below the 1\% level. We will device a strategy where the dominant contribution proportional to $C_{\gamma\gamma}(0)$, including radiative corrections, drops out. Radiative corrections then only have a tiny impact of the coefficients of the subleading terms. 
 
\subsection{Reduction of theoretical uncertainties}
\label{subsec:smart}

If we assume that the Higgs couplings to the electron are close to those predicted by the SM, the CP-odd form factors $F_2^V$ are more than two orders of magnitude smaller than $F_1^V$. For a first discussion we can thus focus on the form factors $F_1^V$. Keeping only the numerically significant terms, we find
\begin{equation}
\begin{aligned}
   F_1^{\Upsilon(1S)} &= 0.022\kappa_W - 0.005\kappa_t - 0.005\kappa_{\gamma\gamma} 
    - (0.017\pm 0.001) \kappa_b + \dots \,, \\
   F_1^{J/\psi} &= - 0.137\kappa_W + 0.030\kappa_t + 0.033\kappa_{\gamma\gamma} 
    + (0.030\pm 0.005) \bar\kappa_c + \dots \,, \\
   F_1^\phi &= 0.206\kappa_W - 0.045\kappa_t - 0.049\kappa_{\gamma\gamma} 
    - (0.015\pm 0.001) \bar\kappa_\phi + \dots \,,
\end{aligned}
\end{equation}
where the last term in each line represents the direct contribution. We have dropped the small imaginary parts of the latter, whose impact is tiny. Replacing $\bar\kappa_\phi\approx\bar\kappa_s\approx 0.019\kappa_s$ and $\bar\kappa_c\approx 0.223\kappa_c$ one obtains equivalent expressions in which the modified Higgs couplings are expressed as corrections to the SM Yukawa couplings. The challenge is to detect the small impact of the direct contributions in the last two cases.

To this end, it is essential to have absolute confidence in the precision with which the indirect contributions can be calculated in the SM, and to be able to subtract these contributions in a reliable way without assuming that the SM is correct. The latter task can be accomplished because the off-shellness of the photon in the $h\to\gamma\gamma^*$ contribution as well as the $h\to\gamma Z^*$ contribution in the third graph in Figure~\ref{fig:diags} are both very small effects. It is therefore possible to eliminate the main dependence of the indirect contributions on the new-physics parameters by considering the following ratio of decay rates:
\begin{equation}\label{wonderful_ratio}
   \frac{\mbox{Br}(h\to V\gamma)}{\mbox{Br}(h\to\gamma\gamma)} 
   = \frac{\Gamma(h\to V\gamma)}{\Gamma(h\to\gamma\gamma)} 
   = \frac{8\pi\alpha^2(m_V)}{\alpha}\,\frac{Q_V^2 f_V^2}{m_V^2}
    \left( 1 - \frac{m_V^2}{m_h^2} \right)^2 
    \frac{\big|1-\Delta_V\big|^2 + \big|r_{\rm CP}-\tilde\Delta_V\big|^2}{1+|r_{\rm CP}|^2} \,.
\end{equation}
Taking such a ratio has the additional advantage that one becomes insensitive to the unknown total width of the Higgs boson, and hence one obtains directly the ratio of branching fractions. One can even go one step further and eliminate the sensitivity to the decay constant $f_V$ by using (\ref{GamVee}) and considering the ratio
\begin{equation}\label{secondratio}
   \frac{m_V}{\Gamma(V\to e^+ e^-)}\,
    \frac{\mbox{Br}(h\to V\gamma)}{\mbox{Br}(h\to\gamma\gamma)} 
   = \frac{6}{\alpha} \left( 1 - \frac{m_V^2}{m_h^2} \right)^2 
    \frac{\big|1-\Delta_V\big|^2 + \big|r_{\rm CP}-\tilde\Delta_V\big|^2}{1+|r_{\rm CP}|^2} \,.
\end{equation}
The only remaining hadronic uncertainties are now contained in the calculation of the reduced form factors $F_V$, which we have collected in Table~\ref{tab:FVnumbers}.

The explicit expressions for the various quantities entering the right-hand side of (\ref{wonderful_ratio}) are $r_{\rm CP}=\tilde C_{\gamma\gamma}(0)/C_{\gamma\gamma}(0)$ and
\begin{equation}\label{DeltaVdef}
\begin{aligned}
   \Delta_V &= - \bar\kappa_V\,\frac{F_V}{C_{\gamma\gamma}(0)}\,
    \frac{\pi m_V v}{\alpha(m_V)\,m_h^2}
    - \frac{C_{\gamma\gamma}(x_V)-C_{\gamma\gamma}(0)}{C_{\gamma\gamma}(0)}
    + \frac{m_V^2}{m_Z^2}\,\frac{v_V}{Q_V s_W^2 c_W^2}\,
    \frac{C_{\gamma Z}(0)}{C_{\gamma\gamma}(0)} \,, \\
   \tilde\Delta_V &= - \bar{\tilde\kappa}_V\,\frac{F_V}{C_{\gamma\gamma}(0)}\,
    \frac{\pi m_V v}{\alpha(m_V)\,m_h^2}
    - \frac{\tilde C_{\gamma\gamma}(x_V)-\tilde C_{\gamma\gamma}(0)}{C_{\gamma\gamma}(0)}
    + \frac{m_V^2}{m_Z^2}\,\frac{v_V}{Q_V (s_W c_W)^2}\,
    \frac{\tilde C_{\gamma Z}(0)}{C_{\gamma\gamma}(0)} \,,
\end{aligned}
\end{equation}
where we work to leading order in the small ratios $m_V^2/m_Z^2$ and $x_V=m_V^2/m_h^2$. Since the individual terms in these expressions are all normalized to $C_{\gamma\gamma}(0)$, it is convenient to define an effective parameter $\kappa_{\gamma\gamma}^{\rm eff}$ by normalizing this coefficient to its SM value. Specifically, we write 
\begin{equation}\label{kgagaeff}
\begin{aligned}
   \kappa_{\gamma\gamma}^{\rm eff}
   &= \frac{C_{\gamma\gamma}(0)\hspace{4.7mm}}{\big[ C_{\gamma\gamma}(0) \big]_{\rm SM}}
    = \Big[ 1.275\kappa_W - 0.282\kappa_t + (0.004-0.003i) \kappa_\tau
     + (0.002-0.002i) \kappa_b \\[-1.5mm]
   &\hspace{3.23cm}\mbox{}+ (0.004-0.003i) \bar\kappa_c - 0.306\kappa_{\gamma\gamma}
    \Big]/(1 - 0.006i) \,.
\end{aligned}
\end{equation}
Here and below the omitted contributions have coefficients equal to zero up to the indicated number of digits. From the empirical fact that the Higgs couplings to $W$ bosons and to the third-generation fermions agree with their SM values within errors, it follows that $\kappa_{\gamma\gamma}^{\rm eff}$ cannot be too different from its SM value, except perhaps for a possible new-physics contribution parameterized by $\kappa_{\gamma\gamma}$. Note that a tiny imaginary part of $\kappa_{\gamma\gamma}^{\rm eff}$, which can be present for non-standard values of the $\kappa_i$ parameters, would have no noticeable impact on our analysis. The ratio $r_{\rm CP}$ entering in (\ref{wonderful_ratio}) vanishes in the SM and is entirely due to the various CP-odd new-physics parameters $\tilde\kappa_i$. We find
\begin{equation}
\begin{aligned}
   r_{\rm CP} 
   &= \frac{-(0.429+0.003i) \tilde\kappa_t + (0.004-0.003i) \tilde\kappa_\tau
           + (0.002-0.002i) \tilde\kappa_b}{\kappa_{\gamma\gamma}^{\rm eff}} \\
   &\quad\mbox{}+ \frac{(0.005-0.003i) \bar{\tilde\kappa}_c 
           - (0.306+0.002i) \tilde\kappa_{\gamma\gamma}}{\kappa_{\gamma\gamma}^{\rm eff}} \,.
\end{aligned}
\end{equation}
Under the assumption that the Higgs couplings to the electron are approximately SM like, the upper bounds $|\tilde\kappa_t|<0.01$ and $|\tilde\kappa_{\gamma\gamma}|<0.006$ mentioned above relation (\ref{derela}) imply that the first and last term in the numerator, which have the largest coefficients by far, are at most 0.004 and 0.002 in magnitude, respectively. It follows that the magnitude of $r_{\rm CP}$ can at most be of order 1\%, and hence the impact of this parameter in (\ref{DeltaVdef}) is likely to be negligible. We emphasize that the parameter $r_{\rm CP}$ can in principle be probed experimentally by studying $h\to\gamma\gamma$ decays, in which both photons undergo nuclear conversion to electron-positron pairs \cite{Bishara:2013vya}. In practice, such a measurement appears to be very challenging.
 
We now present our numerical results for the CP-even coefficients $\Delta_V$ for the various mesons. The complete expressions are collected in Appendix~\ref{app:CPodd}. They contain direct contributions proportional to the relevant $\bar\kappa_q$ parameters and indirect contributions, which are due to the power-suppressed $h\to\gamma Z^*\to\gamma V$ contribution and the effect of the off-shellness of the photon in the $h\to\gamma\gamma^*\to\gamma V$ contribution. These latter terms are significantly smaller than the theoretical uncertainties in the direct terms. In our phenomenological analysis we will keep these small effects but evaluate them in the SM. This gives rise to the expressions
\begin{equation}\label{DeltaVresults}
\begin{aligned}
   \Delta_{\rho^0}
   &= \Big[ (0.068\pm 0.006) + i(0.011\pm 0.002) \Big]\, 
    \frac{\bar\kappa_{\rho^0}}{\kappa_{\gamma\gamma}^{\rm eff}} + 0.00002 \,, \\
   \Delta_\omega
   &= \Big[ (0.068\pm 0.006) + i(0.011\pm 0.002) \Big]\,
    \frac{\bar\kappa_\omega}{\kappa_{\gamma\gamma}^{\rm eff}} - 0.00011 \,, \\
   \Delta_\phi 
   &= \Big[ (0.093\pm 0.008) + i(0.015\pm 0.003) \Big]\,
    \frac{\bar\kappa_\phi}{\kappa_{\gamma\gamma}^{\rm eff}} + 0.00014  \,, \\
   \Delta_{J/\psi} 
   &= \Big[ (0.281\pm 0.045) + i(0.040\pm 0.009) \Big]\,
    \frac{\bar\kappa_c}{\kappa_{\gamma\gamma}^{\rm eff}} + 0.00005 \,,
\end{aligned}
\end{equation}
and
\begin{equation}\label{DeltaVresults2}
\begin{aligned}
   \Delta_{\Upsilon(1S)}  
   &= \Big[ (0.948\pm 0.040) + i(0.130\pm 0.019) \Big]\,
    \frac{\kappa_b}{\kappa_{\gamma\gamma}^{\rm eff}} + 0.0184 - 0.0015 i \,, \\    
   \Delta_{\Upsilon(2S)}  
   &= \big[ (1.014\pm 0.054) + i(0.141\pm 0.022) \big]\,
    \frac{\kappa_b}{\kappa_{\gamma\gamma}^{\rm eff}} + 0.0207 - 0.0015 i \,, \\
   \Delta_{\Upsilon(3S)}  
   &= \Big[ (1.052\pm 0.060) + i(0.148\pm 0.025) \Big]\,
    \frac{\kappa_b}{\kappa_{\gamma\gamma}^{\rm eff}} + 0.0221 - 0.0015 i \,.
\end{aligned}
\end{equation}
Approximate expressions for $\bar\kappa_{\rho^0}$, $\bar\kappa_\omega$ and $\bar\kappa_\phi$ have been given in (\ref{effectivecharges}) and (\ref{rhoomegamixed}). The constant terms in the above results show the tiny power-suppressed corrections. Only for the $\Upsilon(nS)$ states they reach the level of percent. Our complete expressions for the CP-odd coefficients $\tilde\Delta_V$ are also given in Appendix~\ref{app:CPodd}. It is a good approximation to only keep the direct contributions in these terms, which are likely to give rise to the dominant effects. Their coefficients are the same as in the expressions above, but with $\bar\kappa_q$ replaced by $\bar{\tilde\kappa}_q$ and $\kappa_b$ replaced by $\tilde\kappa_b$.

It is interesting to compare our result for the quantities $\Delta_V$ with corresponding expressions obtained by other authors. From \cite{Kagan:2014ila} one can extract $\Delta_{\rho^0}=(0.095\pm 0.020)\,(2\bar\kappa_u+\bar\kappa_d)/3$, $\Delta_\omega=(0.092\pm 0.021)\,(2\bar\kappa_u+\bar\kappa_d)$ and $\Delta_\phi=(0.130\pm 0.027) \bar\kappa_s$, while from \cite{Bodwin:2014bpa} one can obtain $\Delta_{J/\psi}=(0.392\pm 0.053)\bar\kappa_c$, $\Delta_{\Upsilon(1S)}=(1.048\pm 0.046)\kappa_b$, $\Delta_{\Upsilon(2S)}=(1.138\pm 0.053)\kappa_b$ and $\Delta_{\Upsilon(3S)}=(1.175\pm 0.056)\kappa_b$. These values are systematically higher than ours due to the fact that these authors have not (or not fully) included QCD radiative corrections and RG evolution effects in the direct contributions. For the $\Upsilon(nS)$ states it is important to keep the small imaginary parts of the direct contributions, since in the SM the real parts almost perfectly cancel in the combinations $\big|1-\Delta_V\big|$ in (\ref{wonderful_ratio}). The result for $\Delta_\omega$ obtained in \cite{Kagan:2014ila} misses the contribution from $\omega\!-\!\phi$ mixing and contains a sign mistake in front of $\bar\kappa_d$. Note also that our predictions for the $\Delta_V$ parameters of light mesons are significantly more accurate than those obtained in~\cite{Kagan:2014ila}.

\section{Phenomenological results}
\label{sec:pheno}

We begin by quoting our benchmark results for the $h\to V\gamma$ branching fractions in the SM. For a Higgs mass of $m_h=(125.09\pm 0.024)$~GeV, the SM value of the $h\to\gamma\gamma$ branching ratio is $(2.28\pm 0.11)\cdot 10^{-3}$ \cite{Heinemeyer:2013tqa}. Using this result, we obtain for the decays into light vector mesons
\begin{equation}
\begin{aligned}
   \mbox{Br}(h\to\rho^0\gamma) 
   &= (1.68\pm 0.02_{f_\rho}\pm 0.08_{h\to\gamma\gamma})\cdot 10^{-5} \,, \\
   \mbox{Br}(h\to\omega\gamma) 
   &= (1.48\pm 0.03_{f_\omega}\pm 0.07_{h\to\gamma\gamma})\cdot 10^{-6} \,, \\
   \mbox{Br}(h\to\phi\gamma) 
   &= (2.31\pm 0.03_{f_\phi}\pm 0.11_{h\to\gamma\gamma})\cdot 10^{-6} \,, 
\end{aligned}
\end{equation}
where we quote separately the uncertainties due to the vector-meson decay constant $f_V$ and the $h\to\gamma\gamma$ branching ratio, the latter being the dominant source of uncertainty. Our predictions are systematically lower and more accurate than those obtained in \cite{Kagan:2014ila}, where the values $\mbox{Br}(h\to\rho^0\gamma)=(1.9\pm 0.15)\cdot 10^{-5}$, $\mbox{Br}(h\to\omega\gamma)=(1.6\pm 0.17)\cdot 10^{-6}$ and $\mbox{Br}(h\to\phi\gamma)=(3.0\pm 0.13)\cdot 10^{-6}$ are quoted. While the first two results are compatible with ours within errors, there is a significant difference  for the important mode $h\to\phi\gamma$. For decays into heavy vector mesons, we find
\begin{equation}\label{heavyBRs}
\begin{aligned}
   \mbox{Br}(h\to J/\psi\,\gamma) 
   &= (2.95\pm 0.07_{f_{J/\psi}}\pm 0.06_{\rm direct}\pm 0.14_{h\to\gamma\gamma})
    \cdot 10^{-6} \,, \\
   \mbox{Br}(h\to\Upsilon(1S)\,\gamma) 
   &= (4.61\pm 0.06_{f_{\Upsilon(1S)}}\,_{-\,1.21}^{+\,1.75}\,_{\rm direct}
    \pm 0.22_{h\to\gamma\gamma}) \cdot 10^{-9} \,, \\
   \mbox{Br}(h\to\Upsilon(2S)\,\gamma) 
   &= (2.34\pm 0.04_{f_{\Upsilon(2S)}}\,_{-\,0.99}^{+\,0.75}\,_{\rm direct}
    \pm 0.11_{h\to\gamma\gamma}) \cdot 10^{-9} \,, \\
   \mbox{Br}(h\to\Upsilon(3S)\,\gamma) 
   &= (2.13\pm 0.04_{f_{\Upsilon(3S)}}\,_{-\,1.12}^{+\,0.75}\,_{\rm direct}
    \pm 0.10_{h\to\gamma\gamma}) \cdot 10^{-9} \,.
\end{aligned}
\end{equation}
In these cases there is an extra source of theoretical uncertainty related to the calculation of the direct contribution to the decay amplitude. Note that there is an almost perfect cancellation between the direct and indirect contributions to the $h\to\Upsilon(nS)\,\gamma$ decay amplitudes, and as a consequence the resulting branching ratios are roughly three orders of magnitude smaller than the $h\to J/\psi\,\gamma$ branching fraction. For comparison, we note that the branching ratios found in \cite{Bodwin:2014bpa} read $(2.79\,_{-\,0.15}^{+\,0.16})\cdot 10^{-6}$ for $J/\psi$, $(0.61\,_{-\,0.61}^{+\,1.74})\cdot 10^{-9}$ for $\Upsilon(1S)$, $(2.02\,_{-\,1.28}^{+\,1.86})\cdot 10^{-9}$ for $\Upsilon(2S)$ and $(2.44\,_{-\,1.30}^{+\,1.75})\cdot 10^{-9}$ for $\Upsilon(3S)$. We find good agreement with the results reported by these authors except for the decay $h\to\Upsilon(1S)\,\gamma$, where their value is about a factor~7 smaller than ours. The reason is that we do not neglect the imaginary part of the direct contribution to $\Delta_{\Upsilon(1S)}$ in (\ref{DeltaVresults}), which prevents $\left|1-\Delta_{\Upsilon(1S)}\right|^2$ from becoming arbitrarily small. 

Our predictions may also be compared with the upper limits obtained from a recent first analysis of these rare decays reported by the ATLAS collaboration. They are $\mbox{Br}(h\to J/\psi\,\gamma)<1.5\cdot 10^{-3}$, $\mbox{Br}(h\to\Upsilon(1S)\,\gamma)<1.3\cdot 10^{-3}$, $\mbox{Br}(h\to\Upsilon(2S)\,\gamma)<1.9\cdot 10^{-3}$ and $\mbox{Br}(h\to\Upsilon(3S)\,\gamma)<1.3\cdot 10^{-3}$, all at 95\% CL \cite{Aad:2015sda}. It will require an improvement by a factor~500 to become sensitive to the $h\to J/\psi\,\gamma$ mode in the SM, while the SM branching fractions for the decays $h\to\Upsilon(nS)\,\gamma$ are out of reach at the LHC. Nevertheless, as we will discuss below, these decay modes allow for very interesting new-physics searches. With 3\,ab$^{-1}$ of integrated luminosity, about $1.7\times 10^8$ Higgs bosons per experiment will have been produced by the end of the high-luminosity LHC run \cite{Dawson:2013bba}. If the $J/\psi$ is reconstructed via its leptonic decays into muon pairs, the effective branching ratio in the SM is $\mbox{Br}(h\to J/\psi\,\gamma\to\mu^+\mu^-\gamma)=1.8\cdot 10^{-7}$, meaning that about 30 events can be expected per experiment. If also the decays into $e^+ e^-$ can be used, then ATLAS and CMS can hope to collect a combined sample of about 120 events. A detailed discussion of the experimental prospects for reconstructing these events over the background can be found in \cite{Bodwin:2013gca}. Concerning the $h\to\phi\gamma$ decay mode, a reconstruction efficiency $\epsilon_{\phi\gamma}=0.75$ was assumed for the $\phi\gamma$ final state in \cite{Kagan:2014ila}, which appears to us as an optimistic assumption. In the SM one expects about $400\epsilon_{\phi\gamma}$ events per experiment in this mode, meaning that the two experiments can hope to look at a combined sample of several hundred events. Likewise, in the SM one expects about $2900\epsilon_{\rho^0\gamma}$ events per experiment in the decay mode $h\to\rho^0\gamma$.

\begin{figure}
\begin{center}
\includegraphics[width=0.31\textwidth]{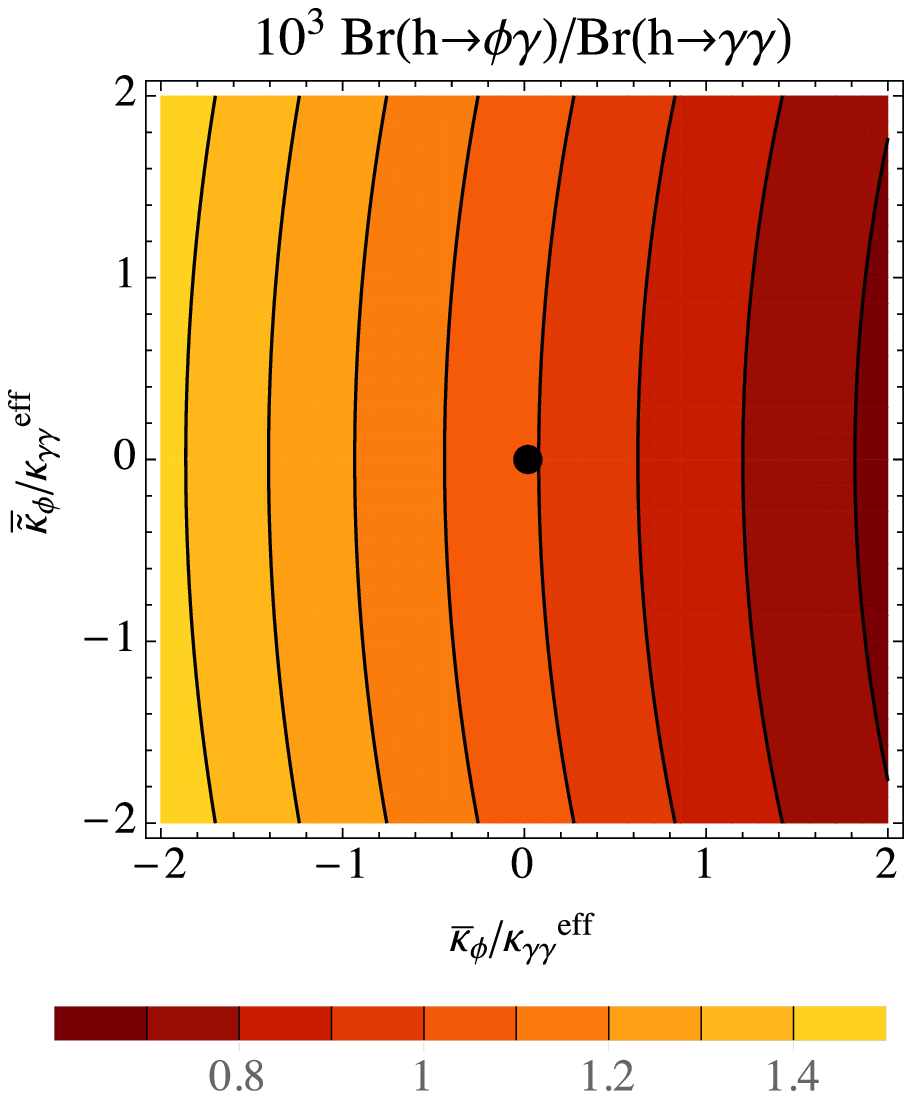} ~ 
\includegraphics[width=0.31\textwidth]{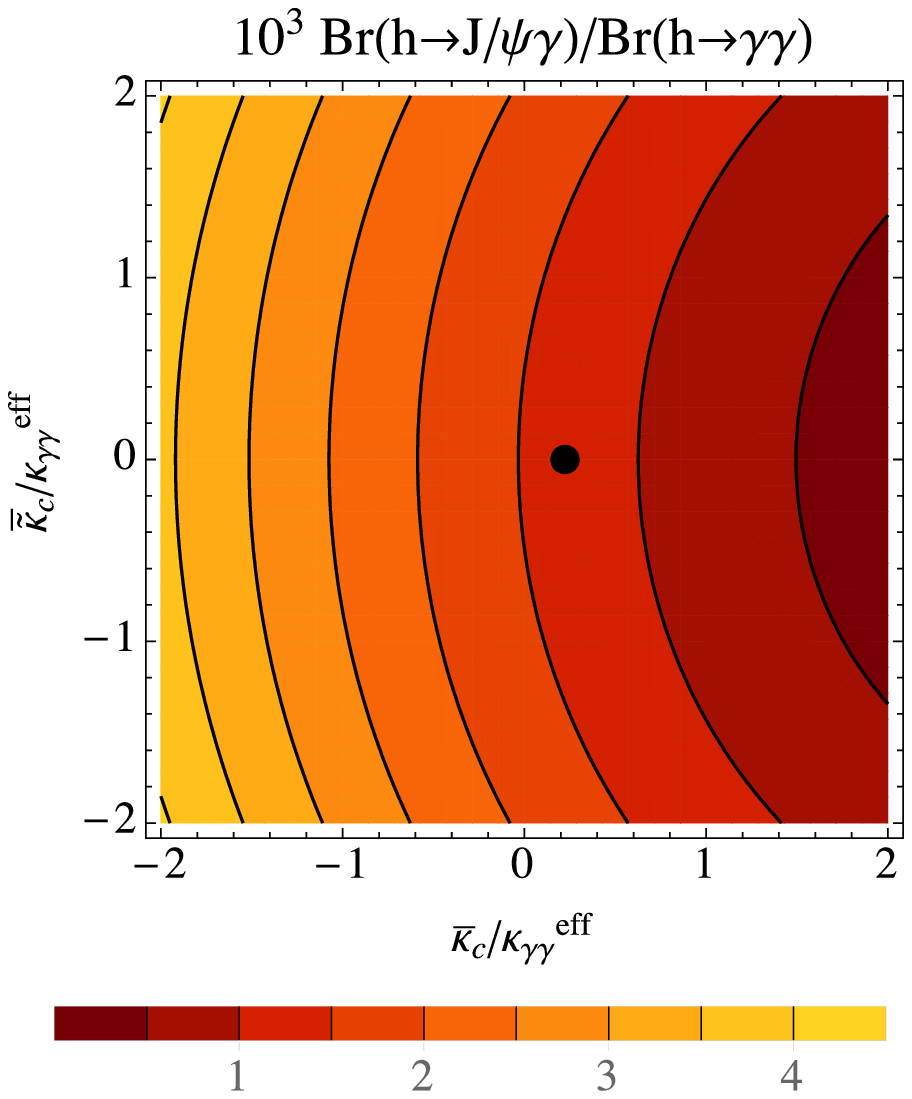} ~
\includegraphics[width=0.31\textwidth]{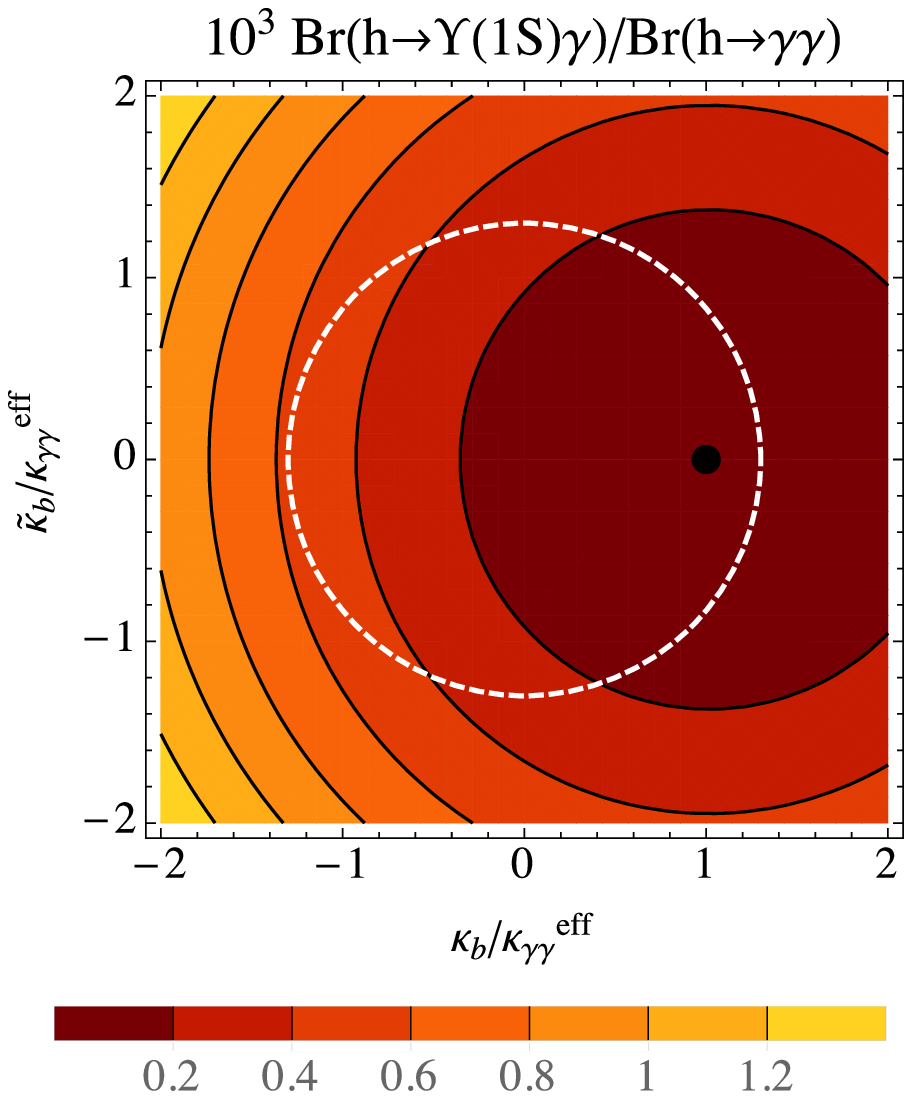}
\parbox{15.5cm}
{\caption{\label{fig:contours}
Predictions (central values) for the ratios of the $h\to V\gamma$ and $h\to\gamma\gamma$ branching fractions with $V=\phi$, $J/\psi$ and $\Upsilon(1S)$ as functions of the rescaled Yukawa couplings normalized to the parameter $\kappa_{\gamma\gamma}^{\rm eff}$ defined in (\ref{kgagaeff}). The black dots indicate the SM values. Coupling parameters inside the dashed white circle in the third plot are preferred by the current LHC data. See text for further details.}}
\end{center}
\end{figure}

In Figure~\ref{fig:contours} we show our predictions for the ratio of branching fractions (times 1000) defined in (\ref{wonderful_ratio}) in the plane of the parameters $\bar\kappa_V/\kappa_{\gamma\gamma}^{\rm eff}$ and $\bar{\tilde\kappa}_V/\kappa_{\gamma\gamma}^{\rm eff}$. We focus on the most interesting cases $V=\phi$, $J/\psi$ and $\Upsilon(1S)$. The corresponding plots for $V=\rho^0$, $\omega$ would look very similar to that for $V=\phi$ (apart from the overall scale of the branching fractions), while the plots for higher $\Upsilon(nS)$ resonances would look very similar to that for the $\Upsilon(1S)$ meson. For orientation, we mention that a value of 0.4 in these plots corresponds to a $h\to V\gamma$ branching fraction of about $10^{-6}$, assuming that the $h\to\gamma\gamma$ branching fraction is SM like. This assumption will be implicit whenever we quote absolute branching ratios below; otherwise the quoted numbers must be rescaled by $\mbox{Br}(h\to\gamma\gamma)/\mbox{Br}(h\to\gamma\gamma)_{\rm SM}$. The structure of our results (\ref{DeltaVresults}) implies that the rescaled Yukawa couplings always enter normalized to $\kappa_{\gamma\gamma}^{\rm eff}$. Hence, if a deviation from the SM is observed in any of these modes, then this could be caused either by a new-physics effect on the $h\to\gamma\gamma$ branching ratio (parameterized by $\kappa_{\gamma\gamma}^{\rm eff}$) or by a non-standard Yukawa coupling. The former effect would however be correlated among all decay channels. We observe that the $h\to\phi\gamma$ branching ratio is rather insensitive to the CP-odd parameter $\bar{\tilde\kappa}_\phi$, the reason being that this parameter enters quadratically and only via the direct contribution, which by itself is a small correction. An analogous statement holds (with less accuracy) for the case $h\to J/\psi\,\gamma$. It is thus a reasonable approximation to study these decay modes under the assumptions that $\bar{\tilde\kappa}_\phi=0$ and $\bar{\tilde\kappa}_c=0$. The situation is different for the $h\to\Upsilon(nS)\,\gamma$ decay modes, for which there is a strong cancellation between the direct and indirect contributions. The direct contributions are no longer a small correction, and hence the quadratic terms in $\kappa_b$ and $\tilde\kappa_b$ are important. The dashed white circle in the third plot indicates the current upper bound on the combination 
\begin{equation}\label{lambdabgdef}
   \lambda_{b\gamma}\equiv
    \sqrt{ \left| \frac{\kappa_b}{\kappa_{\gamma\gamma}^{\rm eff}} \right|^2
          + \left| \frac{\tilde\kappa_b}{\kappa_{\gamma\gamma}^{\rm eff}} \right|^2 } \,.
\end{equation}
To an excellent approximation, $\lambda_{b\gamma}^2$ measures the deviation of the ratio $\mbox{Br}(h\to b\bar b)/\mbox{Br}(h\to\gamma\gamma)$ from its SM value. The Higgs bosons must be produced via the same production mechanism in both cases, so that possible new-physics effects in Higgs production cancel out. Since the $h\to b\bar b$ mode is measured at the LHC in the rare $Vh$ and $t\bar t h$ associated-production channels, at present no accurate direct measurements of $\lambda_{b\gamma}$ are available. However, from the model-independent global analyses of Higgs couplings performed by ATLAS and CMS, in which all couplings to SM particles (including the effective couplings to photons and gluons) are rescaled by corresponding $\kappa_i$ parameters and also invisible Higgs decays are allowed, one obtains $\lambda_{b\gamma}=0.63\pm 0.27$ for CMS (see Figure~17 in \cite{Khachatryan:2014jba}) and $\lambda_{b\gamma}=0.67\pm 0.32$ for ATLAS (see Table~9 in \cite{ATLAS_Higgs}). At 95\% CL this (roughly) implies $\lambda_{b\gamma}<1.3$. Within this allowed region, the $h\to\Upsilon(1S)$ branching ratio varies by more than two orders of magnitude and can take values as large as $1.3\cdot 10^{-6}$. This might be accessible in the high-luminosity run of the LHC. If the $\Upsilon(1S)$ meson is reconstructed via its decays into muon or electron pairs, one could then hope for a sample of about 20 events with 3\,ab$^{-1}$ combining the ATLAS and CMS data sets.

\begin{figure}
\begin{center}
\includegraphics[width=0.45\textwidth]{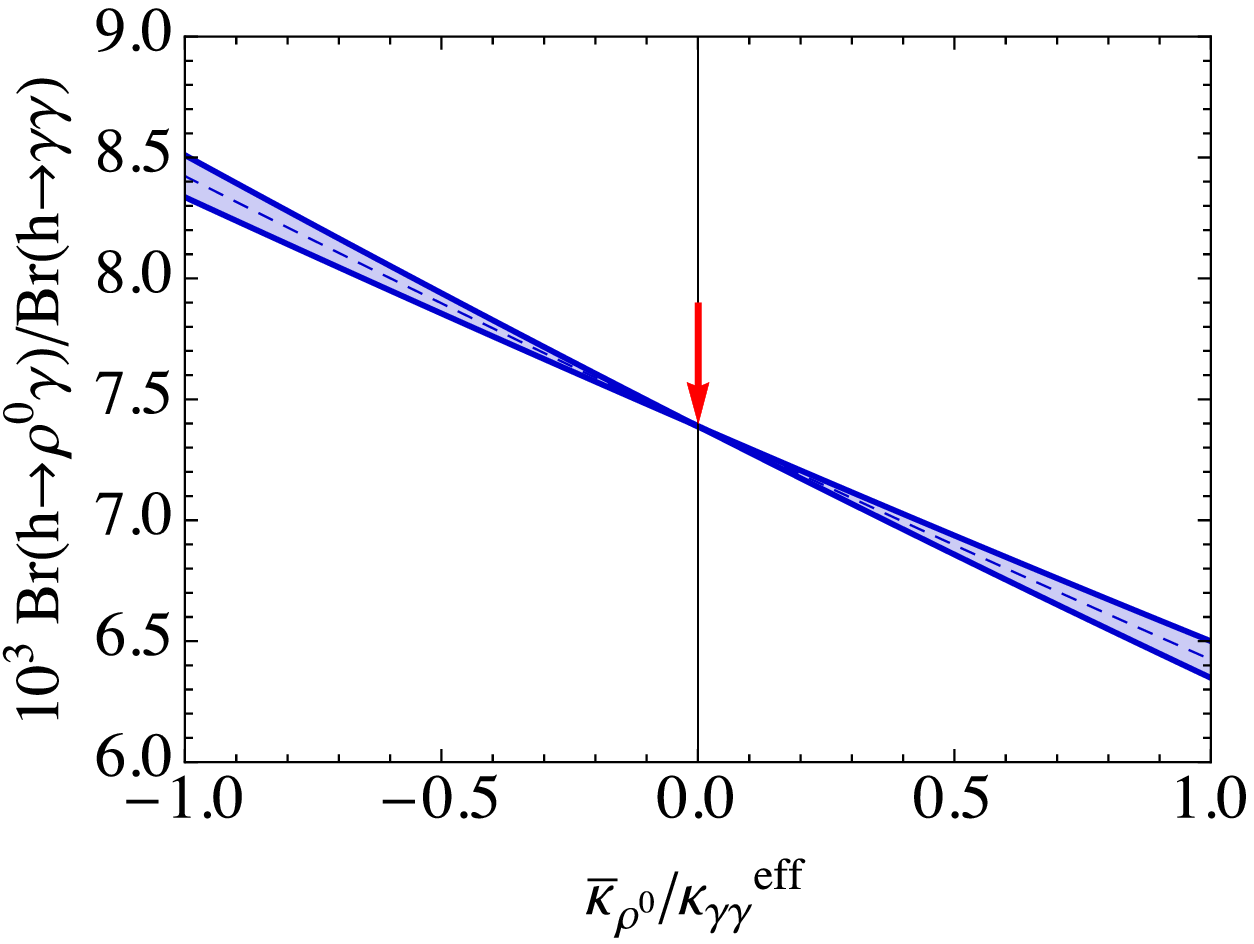} \quad
\includegraphics[width=0.45\textwidth]{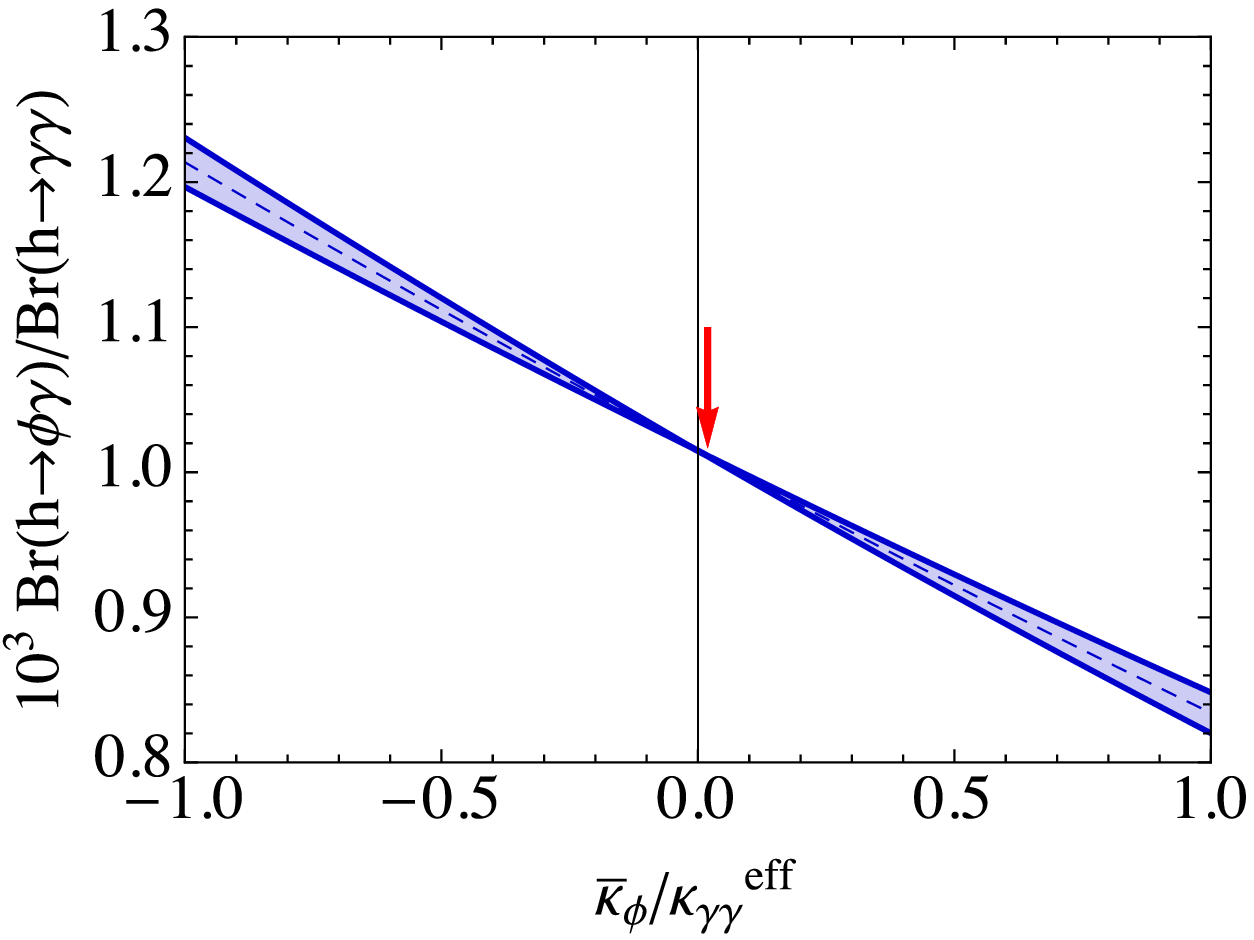}
\parbox{15.5cm}
{\caption{\label{fig:rhophi_ratio}
Predictions for the $h\to\rho\gamma$ and $h\to\phi\gamma$ branching ratios, normalized to the $h\to\gamma\gamma$ branching fraction, as functions of $\bar\kappa_{\rho^0}\approx(2\bar\kappa_u+\bar\kappa_d)/3$ and $\bar\kappa_\phi\approx\bar\kappa_s$, respectively, normalized to $\kappa_{\gamma\gamma}^{\rm eff}$. The SM values are indicated by the red arrows.}}
\end{center}
\end{figure}

In order to better assess the theoretical uncertainties in our predictions, we now study the projections of the results onto the axis where the CP-odd couplings vanish. For the light mesons ($V=\rho^0,\omega,\phi$), setting $\bar{\tilde\kappa}_V=0$ has basically no impact on the branching ratios. In Figure~\ref{fig:rhophi_ratio} we show the ratio of branching fractions defined in (\ref{wonderful_ratio}) as a function of the CP-even couplings $\bar\kappa_V$ for $h\to\rho^0\gamma$ and $h\to\phi\gamma$. The width of the bands indicates the theoretical uncertainty. We have not included the small uncertainty in the values of the decay constants $f_V$, because they can be eliminated using relation (\ref{secondratio}), and we assume that by the time the $h\to V\gamma$ modes will be discovered the decay constants will have been measured more precisely than today. The corresponding plot for $h\to\omega\gamma$ would look identical to the left plot, but with a different vertical scale. While the theoretical uncertainties are small in all cases, we observe that the sensitivity of the branching ratios to the modified Yukawa couplings is unfortunately rather week. For example, a hypothetical 10\% measurement of the $h\to\rho^0\gamma$ branching ratio at the SM value would imply that $|\bar\kappa_{\rho^0}/\kappa_{\gamma\gamma}^{\rm eff}|<0.8$, which is to say that a certain combination of the up- and down-quark Yukawa couplings is bounded not to exceed 80\% of the $b$-quark Yukawa coupling. A 1\% measurement would be required to obtain the more interesting bound $|\bar\kappa_{\rho^0}/\kappa_{\gamma\gamma}^{\rm eff}|<0.08$, which is still more than 100 times the SM value for $\bar\kappa_{\rho^0}$ given in (\ref{effectivecharges}). The situation is not much better for the case $h\to\phi\gamma$. With a 10\% measurement of the branching fraction at the SM rate, one would be able to conclude that $-0.55<\bar\kappa_\phi/\kappa_{\gamma\gamma}^{\rm eff}<0.62$. With a 1\% measurement one would obtain the bounds $-0.04<\bar\kappa_\phi/\kappa_{\gamma\gamma}^{\rm eff}<0.08$, which would come close to the SM value $\bar\kappa_\phi\approx 0.02$. Such a measurement is however out of the reach of the LHC.

\begin{figure}
\begin{center}
\includegraphics[width=0.45\textwidth]{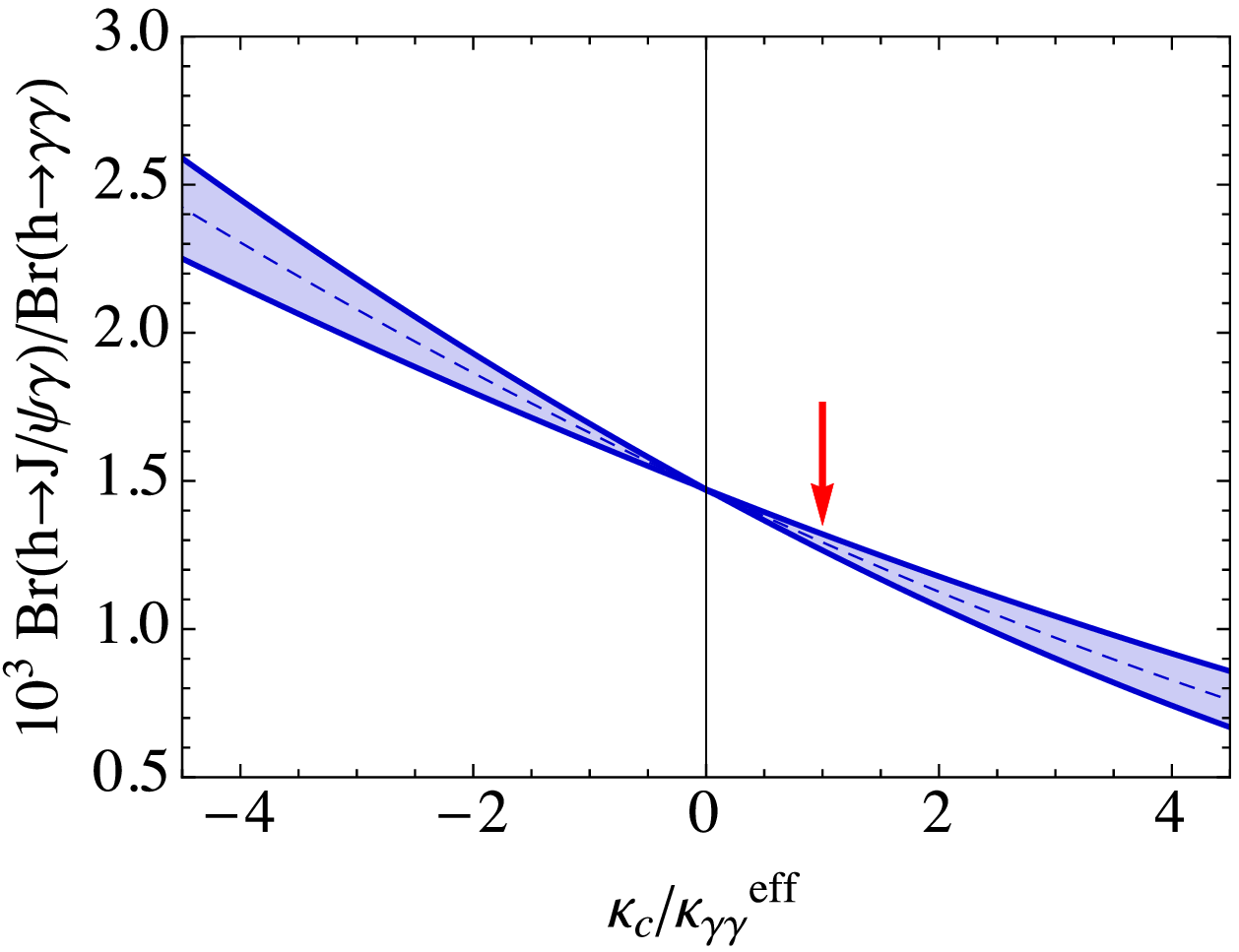} \quad
\includegraphics[width=0.45\textwidth]{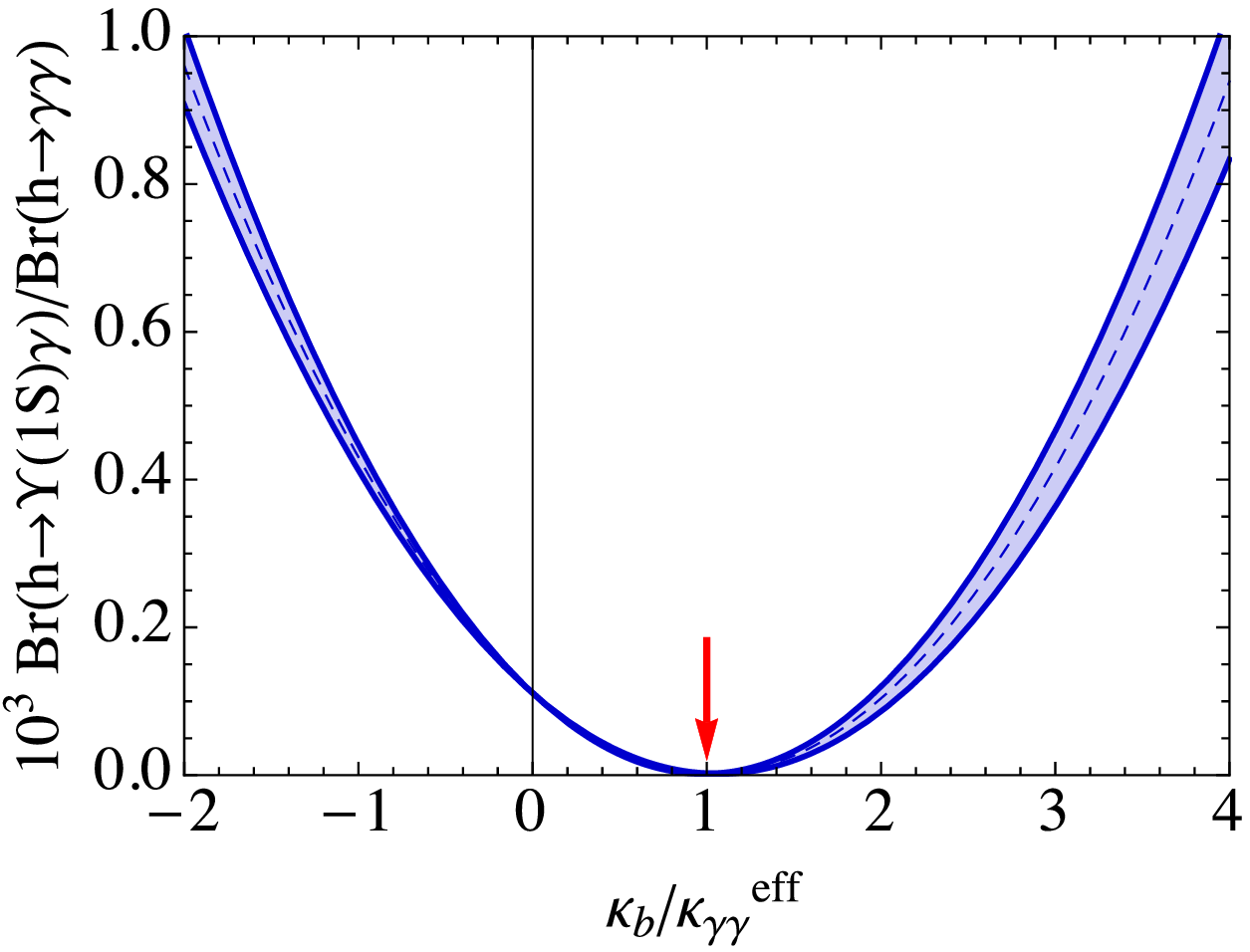}
\parbox{15.5cm}
{\caption{\label{fig:JpsiY1S_ratio}
Predictions for the $h\to J/\psi\,\gamma$ and $h\to\Upsilon(1S)\,\gamma$ branching ratios, normalized to the $h\to\gamma\gamma$ branching fraction, as functions of $\kappa_c$ and $\kappa_b$, respectively, normalized to $\kappa_{\gamma\gamma}^{\rm eff}$. The SM values are indicated by the red arrows.}}
\end{center}
\end{figure}

We now turn to the more interesting cases of radiative Higgs decays into heavy quarkonium states. In Figure~\ref{fig:JpsiY1S_ratio} we show our predictions as a function of the physical parameters $\kappa_c$ (not $\bar\kappa_c$) and $\kappa_b$, again assuming that the CP-odd couplings $\tilde\kappa_c$ and $\tilde\kappa_b$ vanish. In the latter case the impact of a possible CP-odd coupling on the branching fraction can be significant, and in the case of a measurement of a non-standard rate one should keep this possibility in mind. From the left plot in the figure we conclude that a 20\% measurement of the $h\to J/\psi\,\gamma$ branching ratio at the SM value would allow one to constrain $-0.51<\kappa_c/\kappa_{\gamma\gamma}^{\rm eff}<3.07$, which would provide quite interesting information on the CP-even charm-quark Yukawa coupling. With a 10\% measurement this range could be shrunk to $0.32<\kappa_c/\kappa_{\gamma\gamma}^{\rm eff}<1.53$, and with a 5\% measurement one could reach $0.75<\kappa_c/\kappa_{\gamma\gamma}^{\rm eff}<1.19$. Such accurate measurements serve as an interesting physics target for a future 100\,TeV proton-proton collider.

The situation with the $h\to\Upsilon(nS)\,\gamma$ decay modes is different and quite interesting. In the SM the corresponding branching fractions shown in (\ref{heavyBRs}) are so small that these decays would be unobservable. The strong suppression arises from an almost perfect cancellation between the direct and indirect contributions to the decay amplitudes, which results from the fact that in the SM $\mbox{Re}\,\Delta_{\Upsilon(nS)}\approx 1$ within a few percent, see (\ref{DeltaVresults2}). Thanks to this fortuitous fact, these decays offer a much enhanced sensitivity to the effects of new physics. For instance, the SM value of the $h\to\Upsilon(1S)\,\gamma$ branching ratio of $4\times 10^{-9}$ can be enhanced by a factor of more than 200 for $\kappa_b/\kappa_{\gamma\gamma}^{\rm eff}\approx -1$ or $\kappa_b/\kappa_{\gamma\gamma}^{\rm eff}\approx 3$. The first of these possibilities would yield a $h\to b\bar b$ rate consistent with current LHC measurements. For example, with a hypothetical 25\% measurement $\mbox{Br}(h\to\Upsilon(1S)\,\gamma)/\mbox{Br}(h\to\gamma\gamma)=(0.4\pm 0.1)\cdot 10^{-3}$ one would conclude from the figure that $-1.21<\kappa_b/\kappa_{\gamma\gamma}^{\rm eff}<-0.64$, which would be a very significant piece of information and a spectacular sign of new physics. 

One may ask whether the current bounds obtained by the ATLAS collaboration already have a significant impact on the Higgs couplings. Unfortunately this is not the case. We find that the upper values reported in \cite{Aad:2015sda} imply approximately
\begin{equation}
   \sqrt{ \left| \frac{\kappa_c}{\kappa_{\gamma\gamma}^{\rm eff}} \right|^2
    + \left| \frac{\tilde\kappa_c}{\kappa_{\gamma\gamma}^{\rm eff}} \right|^2 } < 429 \,, \qquad
   \sqrt{ \left| \frac{\kappa_b}{\kappa_{\gamma\gamma}^{\rm eff}} \right|^2
    + \left| \frac{\tilde\kappa_b}{\kappa_{\gamma\gamma}^{\rm eff}} \right|^2 } < 78 \,,
\end{equation}
both at 95\% CL. For comparison, we note that $m_t/m_c\approx 268$ and $m_t/m_b\approx 60$. In other words, the current bounds derived from exclusive $h\to V\gamma$ decays imply that the couplings of the charm and bottom quarks to the Higgs boson are not much stronger than the top-quark Yukawa coupling (the more optimistic value $|\kappa_c|<220$ was quoted in \cite{Perez:2015aoa}).   

\begin{figure}
\begin{center}
\includegraphics[width=0.4\textwidth]{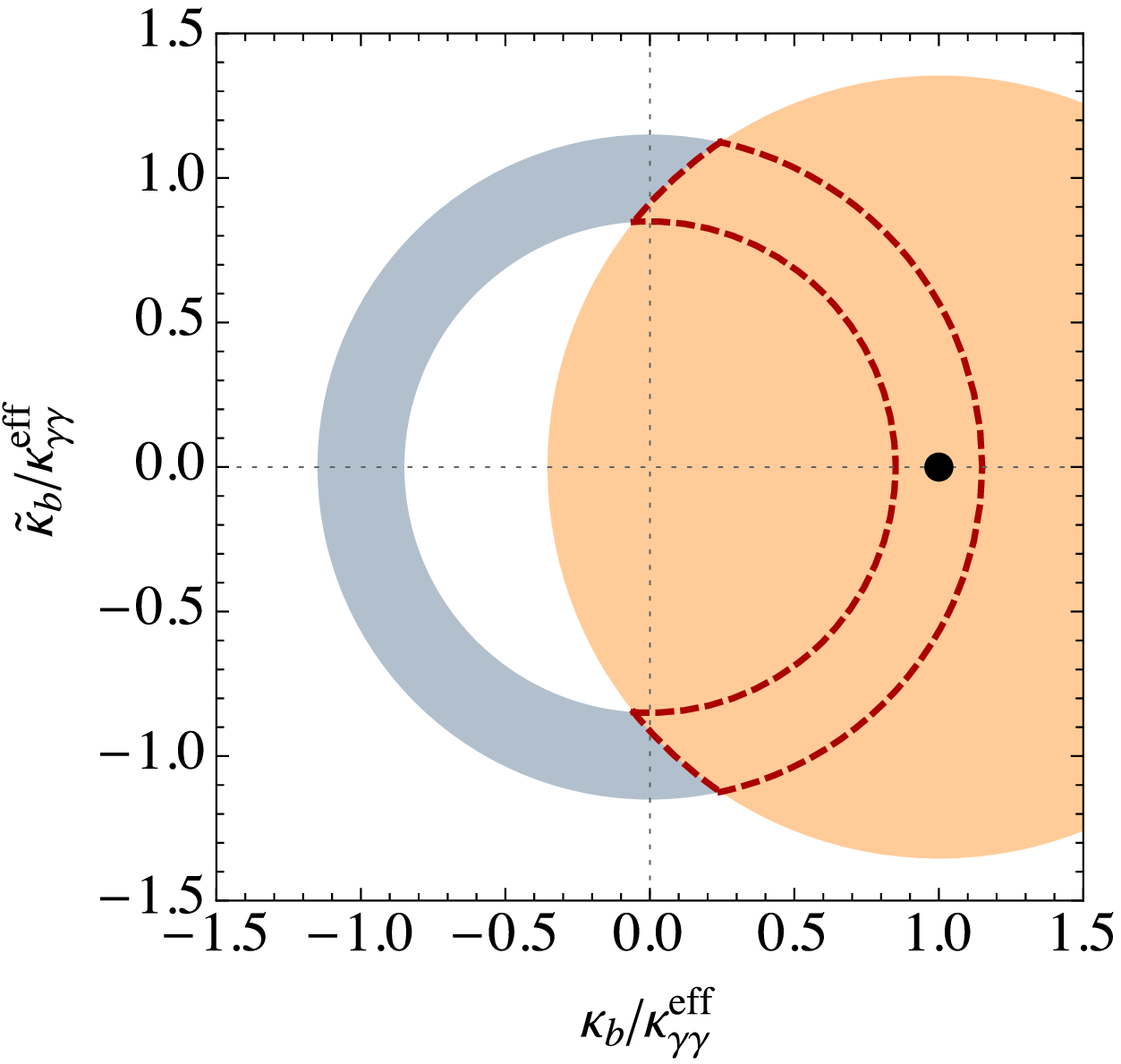} \quad
\includegraphics[width=0.4\textwidth]{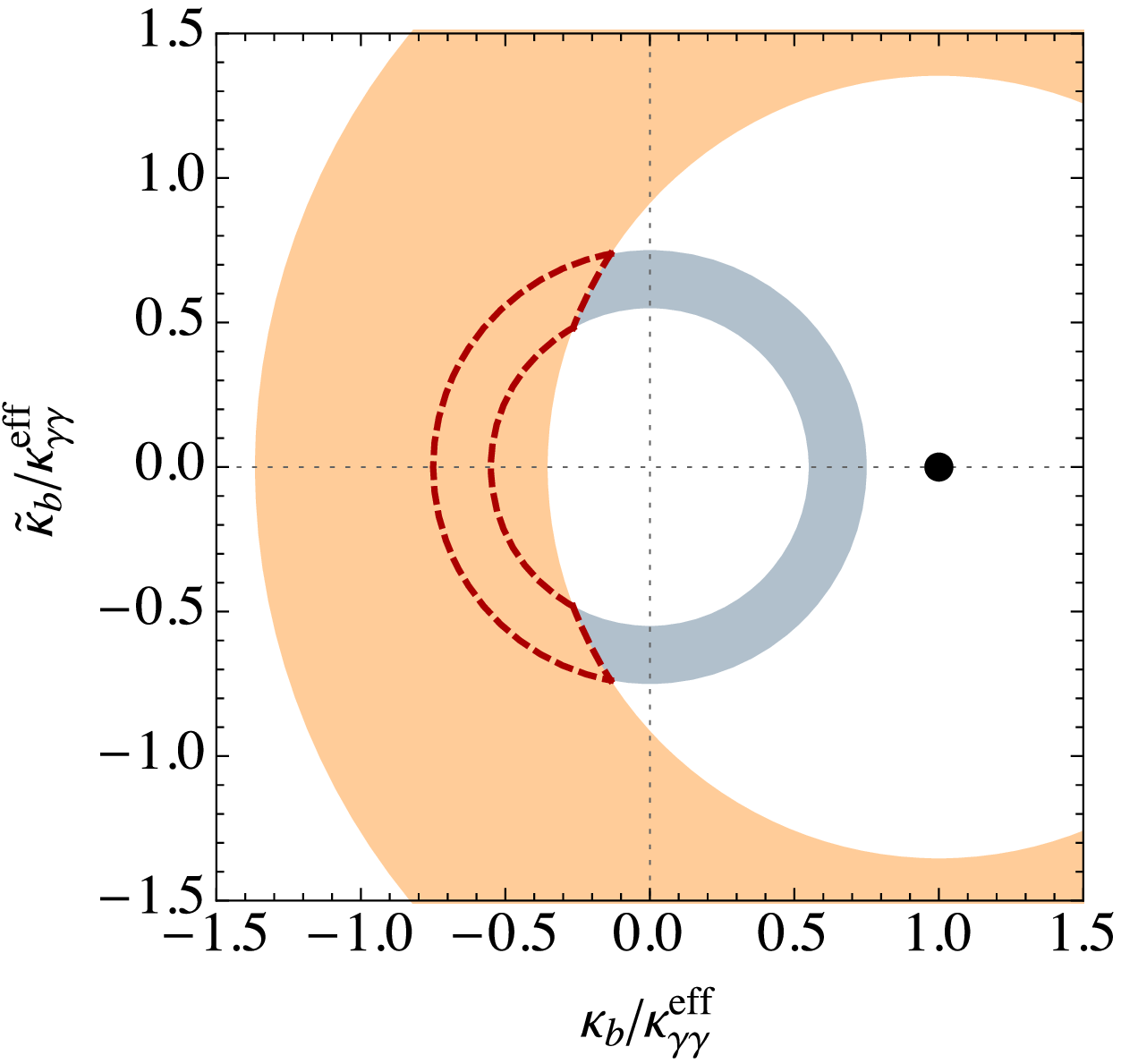}
\parbox{15.5cm}
{\caption{\label{fig:bright_sunshine}
Constraints on the effective coupling strengths $\kappa_b/\kappa_{\gamma\gamma}^{\rm eff}$ and $\tilde\kappa_b/\kappa_{\gamma\gamma}^{\rm eff}$ derived in two possible scenarios for future measurements of the ratios $\mbox{Br}(h\to b\bar b)/\mbox{Br}(h\to\gamma\gamma)$ (light blue) and $\mbox{Br}(h\to\Upsilon(1S)\,\gamma)/\mbox{Br}(h\to\gamma\gamma)$ (orange). The allowed parameter space is given by the red-shaded intersection of the two rings. The black dot indicates the SM value.}}
\end{center}
\end{figure}

We emphasize again that any experimental information on the rare radiative decays $h\to\Upsilon(nS)\,\gamma$ should be interpreted in terms of an allowed region in the two-dimensional plane of the couplings $\kappa_b/\kappa_{\gamma\gamma}^{\rm eff}$ and $\tilde\kappa_b/\kappa_{\gamma\gamma}^{\rm eff}$. The one-dimensional projection shown in Figure~\ref{fig:JpsiY1S_ratio} is meant for illustrative purposes only. It is interesting to speculate about some possible scenarios that may arise at the end of the high-luminosity LHC run with an integrated luminosity of 3\,ab$^{-1}$. Existing estimates of the precision achievable on the Yukawa coupling to bottom quarks and the effective Higgs coupling to photons (our parameter $|\kappa_{\gamma\gamma}^{\rm eff}|$) suggest that, at 95\% CL, the quantity $\lambda_{b\gamma}$ defined in (\ref{lambdabgdef}) can be measured with a precision at least as good as 15\% \cite{Dawson:2013bba}. In Figure~\ref{fig:bright_sunshine} we consider two possible future scenarios:
\begin{equation}
\begin{aligned}
   \mbox{(I)} \qquad
   \lambda_{b\gamma} &= 1.0\pm 0.15 \,, \qquad
   &\frac{\mbox{Br}(h\to\Upsilon(1S)\,\gamma)}{\mbox{Br}(h\to\gamma\gamma)} 
   &< 0.2\cdot 10^{-3} \,, \\
   \mbox{(II)} \qquad
   \lambda_{b\gamma} &= 0.65\pm 0.10 \,, \qquad
   &\frac{\mbox{Br}(h\to\Upsilon(1S)\,\gamma)}{\mbox{Br}(h\to\gamma\gamma)} 
   &= (0.4\pm 0.2)\cdot 10^{-3} \,.
\end{aligned}
\end{equation}
In the first scenario the ratio $\mbox{Br}(h\to b\bar b)/\mbox{Br}(h\to\gamma\gamma)$ is measured at its SM value and an upper limit of about $0.5\cdot 10^{-6}$ is placed on the $h\to\Upsilon(1S)\,\gamma$ branching ratio. Even though no observation of this rare decay is accomplished, the assumed upper bound still has a non-trivial impact, as it limits the allowed values of $\kappa_b/\kappa_{\rm \gamma\gamma}^{\rm eff}$ to the right half-plane. In particular, this would exclude the possibility that $\kappa_b/\kappa_{\rm \gamma\gamma}^{\rm eff}\approx-1$. In the second scenario the parameter $\lambda_{b\gamma}$ is measured close to its current value but with higher accuracy, while a rough 50\% measurement of the $h\to\Upsilon(1S)\,\gamma$ branching ratio at about $(0.91\pm 0.46)\cdot 10^{-6}$ is obtained. In this case one could exclude the SM value $\kappa_b/\kappa_{\rm \gamma\gamma}^{\rm eff}=1$ and limit the allowed values of this ratio to the left half-plane. These speculative results nicely indicate the power of future searches for the rare exclusive decay $h\to\Upsilon(1S)\,\gamma$ in the high-luminosity phase of the LHC. The statistics of such a search can be approximately doubled by including also the $\Upsilon(2S)\,\gamma$ and $\Upsilon(3S)\,\gamma$ final states. 

Several ideas for constraining the absolute value of the charm-quark Yukawa coupling in the high-luminosity run at the LHC or at a future 100\,TeV collider have been put forward in \cite{Perez:2015aoa}. If such a measurement can indeed be made, it implies a circular allowed region centered at $(0,0)$ in the plane of the parameters $\kappa_c/\kappa_{\gamma\gamma}^{\rm eff}$ and $\tilde\kappa_c/\kappa_{\gamma\gamma}^{\rm eff}$. A measurement of the $h\to J/\psi\,\gamma$ branching ratio would be mainly sensitive to $\kappa_c/\kappa_{\gamma\gamma}^{\rm eff}$ and hence confine the couplings to a curved band, which intersects this region (see the center plot in Figure~\ref{fig:contours}). In this way, it may be possible to perform an analysis similar to that shown in Figure~\ref{fig:bright_sunshine} for the Higgs couplings to the charm quark. 

If at a future 100\,TeV proton-proton collider one succeeds to collect large data samples of the rare decays $h\to V\gamma\to l^+ l^-\gamma$ with $V=J/\psi$ or $\Upsilon(nS)$, one might even speculate about the possibility to measure the polarization of both the vector meson and the photon by considering events in which the photon undergoes a nuclear conversion to an electron-positron pair, in analogy with what was proposed for $h\to\gamma\gamma$ decay in \cite{Bishara:2013vya} and for the $B\to K^*\gamma$ process in \cite{Grossman:2000rk}. With such a measurement it would be possible to differentiate between the two structures in (\ref{Ampl}), which in 3-vector notation correspond to the products $\bm{\varepsilon}_V^*\cdot\bm{\varepsilon}_\gamma^*$ and $\bm{\varepsilon}_V^*\times\bm{\varepsilon}_\gamma^*$ of polarization vectors. In this way one would become sensitive to the sign of the ratio $\tilde\kappa_q/\kappa_q$ of the CP-odd and CP-even Yukawa couplings, thus breaking the degeneracy beetween the upper and lower half-planes in Figure~\ref{fig:bright_sunshine}.

\section{Conclusions}
\label{sec:concl}

We have performed a state-of-the-art analysis of the rare exclusive decays $h\to V\gamma$ in the context of generic extensions of the SM with modified Higgs couplings. These decays are characterized by a destructive interference between two decay topologies. The direct contribution is governed by diagrams where the Higgs boson decays into a quark anti-quark pair, from which the vector meson is formed. It is proportional to the (modified) Yukawa coupling of the Higgs to that quark flavor. Using QCD factorization and techniques developed in \cite{Grossman:2015lea}, we have derived a closed analytic expression for the direct amplitudes at next-to-leading order in $\alpha_s$ as an infinite sum over Gegenbauer moments renormalized at the scale $\mu\sim m_h$. In this way large logarithmic corrections are resummed, and the sensitivity of our predictions to hadronic input parameters is reduced. Power corrections to the factorization formula are suppressed by $(\Lambda_{\rm QCD}/m_h)^2$ or $(m_V/m_h)^2$ and can be safely neglected. The second, indirect contribution to the decay amplitude arises from diagrams where the Higgs boson decays into a photon and an off-shell neutral gauge boson, which converts into the vector meson. Due to the smallness of the Yukawa couplings, this indirect contribution is the dominant one in all cases but $h\to\Upsilon(nS)\,\gamma$. In order to reduce the sensitivity to possible new-physics effects in the effective (loop-induced) $h\gamma\gamma$ vertex and thus get access to the quark Yukawa couplings, we consider the ratio of the $h\to V\gamma$ and $h\to\gamma\gamma$ branching fractions, in which the indirect contribution drops out up to very small power corrections. In our analysis we account for the effects of flavor mixing, which can be important for the light mesons $\rho^0$, $\omega$ and~$\phi$.

We have derived numerical predictions for the $h\to V\gamma$ branching fractions and studied the possibility of using such measurements as probes of new-physics effects on the light-quark Yukawa couplings. In the SM, the branching ratios we find typically lie in the range of few times $10^{-6}$, several orders of magnitude below the ATLAS bounds for the $h\to J/\psi\,\gamma$ and $h\to\Upsilon(nS)\,\gamma$ branching ratios reported in \cite{Aad:2015sda}. Nevertheless, the very high yield of Higgs bosons expected in the high-luminosity phase of LHC operation, combined with a dedicated experimental effort, could make measurements of these rare processes possible. We estimate that with 3\,ab$^{-1}$ of integrated luminosity it will be possible to probe for ${\cal O}(1)$ modifications of the real part of the charm-quark Yukawa coupling and ${\cal O}(30)$ modifications of the real part of the strange-quark Yukawa coupling. 

We have emphasized that the decays $h\to\Upsilon(nS)\,\gamma$ provide a golden opportunity to probe for new-physics effects on the bottom-quark Yukawa couplings. Due to a fortuitous cancellation between the direct and indirect contributions to the corresponding decay amplitudes, the SM branching fractions for these modes are more than two orders of magnitude smaller than the $h\to J/\psi\,\gamma$ branching ratio. They are unobservable at the LHC and at any conceivable future collider. However, we show that with ${\cal O}(1)$ modifications of the $b$-quark Yukawa couplings these branching fractions can be enhanced to an observable level. Any measurement of such a decay would be a clear signal of new physics. Moreover, as we have shown, a combined measurement of the two ratios $\mbox{Br}(h\to\Upsilon(nS)\,\gamma)/\mbox{Br}(h\to\gamma\gamma)$ and $\mbox{Br}(h\to b\bar b)/\mbox{Br}(h\to\gamma\gamma)$ can provide complementary information on the real and imaginary parts of the $b$-quark Yukawa coupling. This will allow one to probe the interesting option that $\kappa_b\approx-1$ has the opposite sign as in the SM. 

Several extensions of our work are possible and worth pursuing. Decays into flavor off-diagonal neutral mesons can be used to probe for flavor-violating couplings of the Higgs boson, which at tree level are forbidden in the SM. Another straightforward generalization is the application of our approach to the weak radiative decays $h\to M^0 Z$ and $h\to M^\pm W^\mp$, where $M$ can be a pseudoscalar or vector meson, as well as decays into final states containing two mesons. This is left for future work.

Exclusive radiative decays of the Higgs boson are an experimentally challenging endeavor, because the expected branching fractions are very small and the final states not easy to reconstruct. Nevertheless, we have argued that studies of these processes may not be impossible in the high-luminosity phase at the LHC and, even more so, at a future 100\,TeV proton-proton collider. This would present us with a unique laboratory to study the real and imaginary parts of the Yukawa couplings of bottom and charm quarks, and probe for enhanced Yukawa couplings of the lighter quarks. We cannot think of any other way in which such direct studies of Yukawa couplings could be performed. 

\subsubsection*{Acknowledgments}

We are grateful to J.~Brod, F.~De Fazio, A.~Grozin, U.~Haisch, A.~Kagan, G.~Perez, F.~Yu, J.~Zupan and R.~Zwicky for useful discussions. M.N.\ thanks the Institute for Theoretical Physics in Heidelberg for hospitality and support under a J.~Hans D.~Jensen Professorship. This work has been supported by the Advanced Grant EFT4LHC of the European Research Council (ERC), the Cluster of Excellence {\em Precision Physics, Fundamental Interactions and Structure of Matter\/} (PRISMA -- EXC 1098), grant 05H12UME of the German Federal Ministry for Education and Research (BMBF), and the DFG Graduate School {\em Symmetry Breaking in Fundamental Interactions\/} (GRK 1581).

\begin{appendix}

\section{\boldmath Effects of $\omega\!-\!\phi$ mixing}
\label{app:fVeff}
\renewcommand{\theequation}{A.\arabic{equation}}
\setcounter{equation}{0}

We express the physical mass eigenstates $|\omega\rangle$ and $|\phi\rangle$ in terms of the flavor eigenstates $|\omega_I\rangle=\frac{1}{\sqrt2}\left(|u\bar u\rangle+|d\bar d\rangle\right)$ and $|\phi_I\rangle=|s\bar s\rangle$ by means of the rotation
\begin{equation}
   \left( \begin{array}{c} |\omega\rangle \\ |\phi\rangle \end{array} \right)
   = \left( \begin{array}{cr} \cos\theta_{\omega\phi} &  \; - \sin\theta_{\omega\phi} \\ 
            \sin\theta_{\omega\phi} & \cos\theta_{\omega\phi} \end{array} \right)
   \left( \begin{array}{c} |\omega_I\rangle \\ |\phi_I\rangle \end{array} \right) .
\end{equation}
In the limit where OZI-violating contributions are neglected, we can then relate the matrix elements of the flavor-specific tensor currents in (\ref{vectorme}) to decay constants defined in terms of the analogous matrix elements 
\begin{equation}\label{a.2}
\begin{aligned}
   \langle\omega_I(k,\varepsilon)|\,
    \frac{\bar u\,i\sigma^{\mu\nu} u + \bar d\,i\sigma^{\mu\nu} d}{\sqrt2}\,|0\rangle 
   &= if_{\omega_I}^\perp \left( k^\mu \varepsilon^{*\nu} - k^\nu\varepsilon^{*\mu} \right) , \\
   \langle\phi_I(k,\varepsilon)|\,\bar s\,i\sigma^{\mu\nu} s\,|0\rangle 
   &= if_{\phi_I}^\perp \left( k^\mu \varepsilon^{*\nu} - k^\nu\varepsilon^{*\mu} \right)
\end{aligned}
\end{equation}
of the flavor eigenstates $|\omega_I\rangle$ and $|\phi_I\rangle$ with the corresponding flavor currents. Assuming isospin symmetry, this gives
\begin{equation}\label{a.3}
   f_\omega^{u\perp} = f_\omega^{d\perp} 
    = \frac{\cos\theta_{\omega\phi}}{\sqrt2}\,f_{\omega_I}^\perp \,, \qquad
   f_\omega^{s\perp} = - \sin\theta_{\omega\phi}\,f_{\phi_I}^\perp \,,
\end{equation}
and
\begin{equation}\label{a.4}
   f_\phi^{s\perp} = \cos\theta_{\omega\phi}\,f_{\phi_I}^\perp \,, \qquad
   f_\phi^{u\perp} = f_\phi^{d\perp} = \frac{\sin\theta_{\omega\phi}}{\sqrt2}\,f_{\omega_I}^\perp \,.
\end{equation}
By definition, the quantities $f_\omega^\perp$ and $f_\phi^\perp$ defined below (\ref{vectorme}) are then given by
\begin{equation}
\begin{aligned}
   f_\omega^\perp &= \cos\theta_{\omega\phi}\,f_{\omega_I}^\perp
    + \sqrt2 \sin\theta_{\omega\phi}\,f_{\phi_I}^\perp \,, \\
   f_\phi^\perp &= \cos\theta_{\omega\phi}\,f_{\phi_I}^\perp
    - \frac{\sin\theta_{\omega\phi}}{\sqrt2}\,f_{\omega_I}^\perp \,.
\end{aligned}
\end{equation}
From (\ref{kappaVdef}) it then follows that
\begin{equation}
\begin{aligned}
   \bar\kappa_\omega 
   &= 2\bar\kappa_u - \bar\kappa_d 
    + \frac{\sqrt{2}\,\delta_\omega}{1+\sqrt{2}\,\delta_\omega}
    \left( \bar\kappa_s + \bar\kappa_d - 2\bar\kappa_u \right) , \\
   \bar\kappa_\phi 
   &= \bar\kappa_s + \frac{\delta_\phi}{\sqrt{2}-\delta_\phi}
    \left( \bar\kappa_s + \bar\kappa_d - 2\bar\kappa_u \right) ,  
\end{aligned}
\end{equation}
where
\begin{equation}\label{a.6}
   \delta_\omega = \frac{f_{\phi_I}^\perp}{f_{\omega_I}^\perp}\,\tan\theta_{\omega\phi} \,, \qquad
   \delta_\phi = \frac{f_{\omega_I}^\perp}{f_{\phi_I}^\perp}\,\tan\theta_{\omega\phi} \,.
\end{equation}
In the limit of SU(3) flavor symmetry the ratio $f_{\phi_I}^\perp/f_{\omega_I}^\perp$ can be replaced by~1. Note that in general the mixing angle $\theta_{\omega\phi}$ and the matrix elements in (\ref{a.2}) may depend on the momentum transfer $k^2$. If this is the case, the values of $f_{\omega_I}^\perp$ and $f_{\phi_I}^\perp$ entering in (\ref{a.3}) and (\ref{a.4}) are different. All parameters entering the quantities $\delta_V$ in (\ref{a.6}) must then be evaluated at $k^2=m_V^2$.

\section{RG evolution equations}
\label{app:RGEs}
\renewcommand{\theequation}{B.\arabic{equation}}
\setcounter{equation}{0}

The running quark masses and transverse decay constants obey the RG equations\footnote{We follow the convention that the anomalous dimensions of coupling constants are defined with the opposite sign than those of operators.}
\begin{equation}
   \mu\,\frac{d}{d\mu}\,m_q(\mu) = \gamma^m(\mu)\,m_q(\mu) \,, \qquad
   \mu\,\frac{d}{d\mu}\,f_V^\perp(\mu) = - \gamma^T(\mu)\,f_V^\perp(\mu) \,, 
\end{equation}
where $\gamma^m$ and $\gamma^T$ are the anomalous dimensions of the quark mass and the QCD tensor current in (\ref{vectorme}). At two-loop order these objects were first obtained in \cite{Tarrach:1980up} and \cite{Broadhurst:1994se}, respectively. At NLO in RG-improved perturbation theory, the evolution equations have the solutions
\begin{equation}\label{RGEf}
\begin{aligned}
   m_q(\mu) 
   &= \left( \frac{\alpha_s(\mu)}{\alpha_s(\mu_0)} \right)^{-\frac{\gamma_0^m}{2\beta_0}}
    \left[ 1 - \frac{\gamma_1^m\beta_0-\beta_1\gamma_0^m}{2\beta_0^2}\,
    \frac{\alpha_s(\mu)-\alpha_s(\mu_0)}{4\pi} + \dots \right] m_q(\mu_0) \,, \\
   f_V^\perp(\mu) 
   &= \left( \frac{\alpha_s(\mu)}{\alpha_s(\mu_0)} \right)^{\frac{\gamma_0^T}{2\beta_0}}
    \left[ 1 + \frac{\gamma_1^T\beta_0-\beta_1\gamma_0^T}{2\beta_0^2}\,
    \frac{\alpha_s(\mu)-\alpha_s(\mu_0)}{4\pi} + \dots \right] f_V^\perp(\mu_0) \,.
\end{aligned}
\end{equation}
The relevant one- and two-loop coefficients of the anomalous dimensions read
\begin{equation}
\begin{aligned}
   \gamma_0^m &= - 6 C_F \,, & \quad
    \gamma_1^m &= - 3 C_F^2 - \frac{97}{3}\,C_F C_A + \frac{20}{3}\,C_F T_F n_f \,, \\
   \gamma_0^T &= 2 C_F \,, & \quad
    \gamma_1^T &= - 19 C_F^2 + \frac{257}{9}\,C_F C_A - \frac{52}{9}\,C_F T_F n_f \,,
\end{aligned}
\end{equation}
where $C_F=4/3$, $C_A=3$ and $T_F=1/2$ for SU$_c$(3). Moreover,
\begin{equation}
   \beta_0 = \frac{11}{3}\,C_A - \frac43\,T_F n_f \,, \qquad
   \beta_1 = \frac{34}{3}\,C_A^2 - \frac{20}{3}\,C_A T_F n_f - 4 C_F T_F n_f 
\end{equation}
are the first two coefficients of the QCD $\beta$-function. Above, $\mu_0$ is a low reference scale, at which the quark masses and decay constants are calculated using some non-perturbative approach. 

The evolution of the leading-twist LCDAs at NLO order has been studied in \cite{Dittes:1983dy,Mikhailov:1984ii,Brodsky:1984xk,Mueller:1993hg,Mueller:1994cn}. Starting at two-loop order the scale dependence of $a_n^{V_\perp}(\mu)$ receives contributions proportional to $a_k^{V_\perp}(\mu)$ with $k=0,\dots,n$. Defining the vector $\vec{a}=(1,a_2^{V_\perp},a_4^{V_\perp},\dots)$ containing the even Gegenbauer moments, one finds $\vec{a}(\mu)=\bm{U}(\mu,\mu_0)\,\vec{a}(\mu_0)$, where the evolution matrix $\bm{U}$ has a triangular structure with entries
\begin{equation}
   U_{nk}(\mu,\mu_0) = \left\{ \begin{array}{ll}
    \displaystyle \frac{\alpha_s(\mu)}{4\pi}\,d_n^k(\mu,\mu_0)\,E_n^{\rm LO}(\mu,\mu_0) & ;~ k<n \,, \\[3mm]
    E_n^{\rm NLO}(\mu,\mu_0) & ;~ k=n \,,
   \end{array} \right.
\end{equation}
and $U_{nk}(\mu,\mu_0)=0$ for $k>n$ (with even $k,n\ge 0$). Explicit expressions for $E_n(\mu,\mu_0)$ and $d_n^k(\mu,\mu_0)$ can be found, e.g., in Appendix~A of \cite{Agaev:2010aq}. The relevant two-loop anomalous dimensions for the Gegenbauer moments of a transversely polarized vector meson were calculated in \cite{Vogelsang:1997ak,Hayashigaki:1997dn}. Numerically, we obtain for the evolution from the low reference scale $\mu_0=1$\,GeV up to the high scale $\mu=m_h$
\begin{equation}
   \bm{U}(m_h,\mu_0) = \left( \begin{array}{ccccccc}
    \phantom{-}1 & \phantom{-}0 & \phantom{-}0 & \phantom{-}0 & \phantom{-}0 & \phantom{-}0 & \dotsm \\
    -0.00335 & \phantom{-}0.463 & \phantom{-}0 & \phantom{-}0 & \phantom{-}0 & \phantom{-}0 & \dotsm \\
    \phantom{-}0.00079 & -0.00716 & \phantom{-}0.304 & \phantom{-}0 & \phantom{-}0 & \phantom{-}0 & \dotsm \\
    \phantom{-}0.00076 & -0.00139 & -0.00608 & \phantom{-}0.228 & \phantom{-}0 & \phantom{-}0 & \dotsm \\
    \phantom{-}0.00054 & -0.00019 & -0.00193 & -0.00484 & \phantom{-}0.182 & \phantom{-}0 & \dotsm \\
    \phantom{-}0.00038 & \phantom{-}0.00011 & -0.00069 & -0.00191 & -0.00388 & \phantom{-}0.152 & \dotsm \\
    \phantom{-}\vdots & \phantom{-}\vdots & \phantom{-}\vdots & \phantom{-}\vdots & \phantom{-}\vdots & 
     \phantom{-}\vdots & \ddots
   \end{array} \right) ,
\end{equation}
where we have adjusted the number of light quark flavors when crossing the thresholds at $\mu=m_c$ and $m_b$. Due to the high value of the electroweak scale the mixing effects are strongly suppressed; for example, we obtain $a_4^{V_\perp}(m_h)\approx 0.304 a_4^{V_\perp}(\mu_0)-0.007 a_2^{V_\perp}(\mu_0)+0.0008$. When NLO evolution effects are included, the coefficients of the various terms shown in the second row of (\ref{IVnumerics}) get rescaled by factors of 1.001, 0.994, 0.984, 0.975, 0.969, respectively. Given that all present estimates of the hadronic parameters $a_n^{V_\perp}$ are afflicted with large theoretical uncertainties, it is sufficient for all practical purposes to use the leading-order solution (\ref{anevol}).

\section{Transverse vector-meson decay constants}
\label{app:fVperp}
\renewcommand{\theequation}{C.\arabic{equation}}
\setcounter{equation}{0}

The direct contributions to the $h\to V\gamma$ form factors in (\ref{FVres}) involve the ratio $f_V^\perp(\mu)/f_V$ of the vector-meson decay constants of tensor and vector currents, as defined in (\ref{fVdef}) and (\ref{vectorme}). It is reasonable to assume that non-perturbative evaluations of this ratio are subject to smaller theoretical uncertainties than calculations of the individual decay constants. For light mesons, predictions for the ratio of decay constants have been obtained using light-cone QCD sum rules and lattice QCD. Table~\ref{tab:fVpfV} shows a compilation of available results, normalized at the reference scale $\mu=2$\,GeV. The numbers in boldface show our own combinations of these results.

\begin{table}[t]
\begin{center}
\begin{tabular}{|c|c|cc|}
\hline 
Meson $V$ & $f_V^\perp(2\,{\rm GeV})/f_V$ & Method & Reference \\
\hline 
$\rho$ & $0.76\pm 0.04$ & lattice (unquenched) & \cite{Jansen:2009hr} \\
 & $0.687\pm 0.027$ & lattice (unquenched) & \cite{Allton:2008pn} \\
 & $0.742\pm 0.014$ & lattice (quenched) & \cite{Braun:2003jg} \\
 & $0.72\pm 0.02\,_{-0.00}^{+0.02}$ & lattice (quenched) & \cite{Becirevic:2003pn} \\
 & $0.70\pm 0.04$ & light-cone sum rules & \cite{Ball:2006eu}  \\
 & $\bm{0.72\pm 0.04}$ & our combination & \\
\hline
$\omega$ & $0.707\pm 0.046$ & light-cone sum rules$^\dagger$ & \cite{Ball:2006eu} \\
 & $\bm{0.71\pm 0.05}$ & our combination & \\
\hline
$\phi$ & $0.750\pm 0.008$ & lattice (unquenched) & \cite{Allton:2008pn} \\
 & $0.780\pm 0.008$ & lattice (quenched) & \cite{Braun:2003jg} \\
 & $0.76\pm 0.01$ & lattice (quenched) & \cite{Becirevic:2003pn} \\
 & $0.763\pm 0.041$ & light-cone sum rules$^\dagger$ & \cite{Ball:2006eu} \\
 & $\bm{0.76\pm 0.04}$ & our combination & \\
\hline
\end{tabular}
\parbox{15.5cm}
{\caption{\label{tab:fVpfV} 
Compilation of theoretical predictions for the ratio $f_V^\perp(\mu)/f_V$ at $\mu=2$\,GeV for light vector mesons. Values marked with a dagger are obtained by taking ratios of individual results for the two decay constants. In our combinations we adopt more conservative error estimates than in some of the original references.}}
\end{center}
\vspace{-3mm}
\end{table} 

For heavy quarkonia, the ratio of decay constants can be calculated using NRQCD. Including the leading relativistic corrections \cite{Bodwin:2014bpa} and one-loop QCD effects \cite{Wang:2013ywc}, we obtain
\begin{equation}
   \frac{f_V^\perp(\mu)}{f_V} 
   = \frac{m_V}{2m_Q} \left( 1 - \frac23\,\langle v^2\rangle_V 
    + \frac{C_F\alpha_s(\mu)}{4\pi}\,\ln\frac{m_Q^2}{\mu^2} + \dots \right) ,
\end{equation}
where $m_Q$ is the pole mass of the heavy quark, and the dots denote higher-order terms. Numerical values for the NRQCD matrix element $\langle v^2\rangle_V$ for the $J/\psi$ and $\Upsilon(nS)$ states with $n=1,2,3$ have been obtained from an analysis of the electromagnetic decays $V\to e^+ e^-$ including ${\cal O}(\alpha_s)$ corrections and the leading relativistic effects \cite{Bodwin:2007fz,Chung:2010vz}. The results obtained in this way, along with the adopted values of the pole masses of the heavy quarks, are compiled in Table~\ref{tab:NRQCDpars}. There are significant uncertainties related to the values of the heavy-quark pole masses and the NRQCD matrix elements $\langle v^2\rangle_V$, which in our opinion have been underestimated in these analyses. It is well known that the concept of a pole mass is ill defined beyond perturbation theory and affected by renormalon ambiguities \cite{Bigi:1994em,Beneke:1994sw}. The values used in \cite{Bodwin:2007fz,Chung:2010vz} are ``one-loop pole masses'', but the intrinsic uncertainties in these values are parametrically of order $\Lambda_{\rm QCD}$. Also, as emphasized in \cite{Beneke:2014qea}, the NRQCD expressions for the electromagnetic decay rates $\Gamma(V\to e^+ e^-)$ receive very large perturbative corrections at two- and three-loop order, and this prevents an accurate extraction of the non-perturbative parameters $\langle v^2\rangle_V$. In view of these remarks, the errors assigned on the $b$-quark mass and on the NRQCD matrix elements for the $\Upsilon (nS)$ states seem overly optimistic. In order to be conservative, we increase these errors by a factor of~2. This yields the results shown in Table~\ref{tab:hadronic_inputs2}.

\begin{table}
\begin{center}
\begin{tabular}{|c|ccc|}
\hline 
Meson $V$ & $m_Q$ [GeV] & $\langle v^2\rangle_V$ & $f_V^\perp(2\,{\rm GeV})/f_V$ \\
\hline 
$J/\psi$ & $1.4\pm 0.2$ & $0.225\,_{-\,0.088}^{+\,0.106}$ \cite{Bodwin:2007fz}
 & $0.91\pm 0.14$ \\
$\Upsilon(1S)$ & $4.6\pm 0.1$ & $-0.009\pm 0.003$ \cite{Chung:2010vz} & $1.09\pm 0.02$ \\
$\Upsilon(2S)$ & $4.6\pm 0.1$ & $0.090\pm 0.011$ \cite{Chung:2010vz} & $1.08\pm 0.02$ \\
$\Upsilon(3S)$ & $4.6\pm 0.1$ & $0.155\pm 0.018$ \cite{Chung:2010vz} & $1.07\pm 0.03$ \\
\hline
\end{tabular}
\parbox{15.5cm}
{\caption{\label{tab:NRQCDpars} 
NRQCD parameters for heavy quarkonia extracted from their electronic decay widths $\Gamma(V\to e^+ e^-)$, and resulting values for the ratio of decay constants.}}
\end{center}
\end{table}

\section{Loop functions}
\label{app:loopfunctions}
\renewcommand{\theequation}{D.\arabic{equation}}
\setcounter{equation}{0}

The loop functions describing fermionic and bosonic contributions to the off-shell $h\to\gamma V^*$ decay amplitudes with $V=\gamma,Z$ have been derived first in \cite{Bergstrom:1985hp}. In our notation, they read\footnote{We have corrected a typo in the expression for the function $A_f(\tau,x)$ given in eq.~(4) of \cite{Bergstrom:1985hp}.}
\begin{equation}
\begin{aligned}
   A_f(\tau,x) 
   &= \frac{3\tau}{2(1-x)} \left\{ 1 - \frac{2x}{1-x}\,
    \Big[ g(\tau) - g\big(\tau/x\big) \Big]
    + \left( 1-\frac{\tau}{1-x} \right) \Big[ f(\tau) - f\big(\tau/x\big) \Big] \right\} , \\
   B_f(\tau,x) 
   &= \frac{\tau}{1-x}\,\Big[ f(\tau) - f\big(\tau/x\big) \Big] \,, \\
   A_W^{\gamma\gamma}(\tau,x) 
   &= \frac{2+3\tau}{1-x} \left\{ 1 - \frac{2x}{1-x}\,
    \Big[ g(\tau) - g\big(\tau/x\big) \Big] \right\}
    + \frac{3\tau}{\left(1-x\right)^2} \left( 2 - \tau - \frac{8x}{3} \right)
    \Big[ f(\tau) - f\big(\tau/x\big) \Big] \,, \\
   A_W^{\gamma Z}(\tau,x) 
   &= \frac{1}{1-x} \left[ 1 - 2s_W^2 
    + \left( \frac52 - 3 s_W^2\! \right) \tau \right] \left\{ 1 - \frac{2x}{1-x}\,
    \Big[ g(\tau) - g\big(\tau/x\big) \Big] \right\} \\
   &\quad\mbox{}+ \frac{\tau}{\left(1-x\right)^2}
    \left[ \left( \frac52 - 3 s_W^2\! \right) (2-\tau) 
    - 2x \big( 3-4s_W^2 \big) \right] \Big[ f(\tau) - f\big(\tau/x\big) \Big] \,, 
\end{aligned}
\end{equation}
where
\begin{equation}
   f(\tau) = \left\{ \begin{array}{cl}
    \displaystyle \arcsin^2\frac{1}{\sqrt\tau} & \,; \quad \tau\ge 1 \,, \\[3.5mm]
    \displaystyle - \frac14 \left( \ln\frac{1+\sqrt{1-\tau}}{1-\sqrt{1-\tau}} - i\pi \right)^2
     & \,; \quad \tau < 1 \,,
   \end{array} \right.
\end{equation}
and $g(\tau)=\tau(1-\tau)\,f'(\tau)$. In the limit $\tau\to\infty$ one finds that $A_f(\tau,x)\to 1$, $B_f(\tau,x)\to 1$, $A_W^{\gamma\gamma}(\tau,x)\to 7$ and $A_W^{\gamma Z}(\tau,x)\to\frac{31}{6}-7s_W^2$.

\section{\boldmath Coefficient functions $\Delta_V$ and $\tilde\Delta_V$}
\label{app:CPodd}
\renewcommand{\theequation}{E.\arabic{equation}}
\setcounter{equation}{0}

Here we list the complete expressions for the CP-even coefficients $\Delta_V$ and the CP-odd coefficients $\tilde\Delta_V$ defined in (\ref{DeltaVdef}). For the former ones, we obtain
\begin{equation}
\begin{aligned}
   \Delta_{\rho^0}
   &= \frac{\big[(0.068\pm 0.006) + i(0.011\pm 0.002) \big] \bar\kappa_{\rho^0}
            + 0.0001\kappa_W - 0.0001\bar\kappa_c}{\kappa_{\gamma\gamma}^{\rm eff}} \,, \\
   \Delta_\omega
   &= \frac{\big[(0.068\pm 0.006) + i(0.011\pm 0.002) \big] \bar\kappa_\omega
            - 0.0001\kappa_W - 0.0001\bar\kappa_c}{\kappa_{\gamma\gamma}^{\rm eff}} \,, \\
   \Delta_\phi 
   &= \frac{\big[(0.093\pm 0.008) + i(0.015\pm 0.003) \big] \bar\kappa_\phi
            + 0.0002\kappa_W - 0.0002\bar\kappa_c - 0.0001\kappa_{\gamma Z}}%
           {\kappa_{\gamma\gamma}^{\rm eff}} \,, \\
   \Delta_{J/\psi} 
   &= \frac{\big[(0.281\pm 0.045) + i(0.040\pm 0.009) \big] \bar\kappa_c}%
           {\kappa_{\gamma\gamma}^{\rm eff}} \\
   &\quad\mbox{}+ \frac{0.0004\kappa_W - 0.0003\kappa_\tau - 0.0001\kappa_b
            + 0.0001\bar\kappa_s - 0.0003\kappa_{\gamma Z}}%
           {\kappa_{\gamma\gamma}^{\rm eff}} \,, \\
   \Delta_{\Upsilon(1S)}  
   &= \frac{\big[(0.948\pm 0.040) + i(0.130\pm 0.019) \big] \kappa_b}%
           {\kappa_{\gamma\gamma}^{\rm eff}} \\
   &\quad\mbox{}+ \frac{0.019\kappa_W - 0.001\kappa_t - 0.001i\kappa_\tau 
            + (0.001-0.002i) \bar\kappa_c - 0.010\kappa_{\gamma Z}}%
           {\kappa_{\gamma\gamma}^{\rm eff}} \,,
\end{aligned}
\end{equation}
where the parameters $\bar\kappa_V$ for light mesons have been defined in (\ref{kappaVdef}). 
The corresponding expressions for the $\Upsilon(2S)$ and $\Upsilon(3S)$ mesons will not be shown explicitly, since they are very similar to that for the $\Upsilon(1S)$ state, see  (\ref{DeltaVresults}). In the above results the first term shows the direct contribution. The remaining terms, which originate from the power-suppressed $h\to\gamma Z^*\to\gamma V$ contribution and the effect of the off-shellness of the photon in the $h\to\gamma\gamma^*\to\gamma V$ contribution, are extremely small. Even for $\Delta_\phi$ and assuming a SM-like Higgs couplings to strange quarks, the theoretical uncertainty in the direct contribution is an order of magnitude larger than the power-suppressed terms. Only for $\Upsilon(1S)$ the power-suppressed terms reach the level of $10^{-2}$, but still this contribution is smaller than the theoretical uncertainty in the direct contribution. Our complete expressions for the CP-odd coefficients $\tilde\Delta_V$ are 
{\small
\begin{equation}
\begin{aligned}
   \tilde\Delta_{\rho^0} - r_{\rm CP}
   &= \frac{\big[(0.068\pm 0.006) + i(0.011\pm 0.002) \big] \bar{\tilde\kappa}_{\rho^0}
            + (0.429+0.003i) \tilde\kappa_t + (0.306+0.002i) \tilde\kappa_{\gamma\gamma}}%
           {\kappa_{\gamma\gamma}^{\rm eff}} \\
   &\quad\mbox{}- \frac{(0.004-0.003i) \tilde\kappa_\tau
            + (0.002-0.002i) \tilde\kappa_b + (0.005-0.003i) \bar{\tilde\kappa}_c}%
           {\kappa_{\gamma\gamma}^{\rm eff}} \,, \\
   \tilde\Delta_\omega - r_{\rm CP} 
   &= \frac{\big[(0.068\pm 0.006) + i(0.011\pm 0.002) \big] \bar{\tilde\kappa}_\omega
            + (0.429+0.003i) \tilde\kappa_t + (0.306+0.002i) \tilde\kappa_{\gamma\gamma}}%
           {\kappa_{\gamma\gamma}^{\rm eff}} \\
   &\quad\mbox{}- \frac{(0.004-0.003i) \tilde\kappa_\tau
            + (0.002-0.002i) \tilde\kappa_b + (0.005-0.003i) \bar{\tilde\kappa}_c}%
           {\kappa_{\gamma\gamma}^{\rm eff}} \,, \\
   \tilde\Delta_\phi - r_{\rm CP}
   &= \frac{\big[(0.093\pm 0.008) + i(0.015\pm 0.003) \big] \bar{\tilde\kappa}_\phi
            + (0.429+0.003i) \tilde\kappa_t + (0.306+0.002i) \tilde\kappa_{\gamma\gamma}}%
           {\kappa_{\gamma\gamma}^{\rm eff}} \\
   &\quad\mbox{}- \frac{(0.004-0.003i) \tilde\kappa_\tau
            + (0.002-0.002i) \tilde\kappa_b + (0.005-0.003i) \bar{\tilde\kappa}_c}%
           {\kappa_{\gamma\gamma}^{\rm eff}} \,, \\
   \tilde\Delta_{J/\psi} - r_{\rm CP}
   &= \frac{\big[(0.277\pm 0.045) + i(0.043\pm 0.009) \big] \bar{\tilde\kappa}_c
            + (0.429+0.003i) \tilde\kappa_t + (0.306+0.002i) \tilde\kappa_{\gamma\gamma}}%
           {\kappa_{\gamma\gamma}^{\rm eff}} \\
   &\quad\mbox{}- \frac{(0.004-0.003i) \tilde\kappa_\tau
            + (0.003-0.002i) \tilde\kappa_b}%
           {\kappa_{\gamma\gamma}^{\rm eff}} \,, \\
   \tilde\Delta_{\Upsilon(1S)} - r_{\rm CP}
   &= \frac{\big[(0.945\pm 0.040) + i(0.132\pm 0.019) \big] \tilde\kappa_b
            + (0.427+0.003i) \tilde\kappa_t + (0.306+0.002i) \tilde\kappa_{\gamma\gamma}}%
           {\kappa_{\gamma\gamma}^{\rm eff}} \\
   &\quad\mbox{}- \frac{(0.004-0.002i) \tilde\kappa_\tau
            + (0.004-0.001i) \bar{\tilde\kappa}_c + 0.010\tilde\kappa_{\gamma Z}}%
           {\kappa_{\gamma\gamma}^{\rm eff}} \,.
\end{aligned}
\end{equation}
}\noindent 
It is a good approximation to only keep the direct contributions, which are likely to give rise to the dominant effects. 

\end{appendix}

\end{document}